\documentclass{ieeeaccess}
\usepackage{cite}
\usepackage{amsmath,amssymb,amsfonts}
\usepackage{algorithmic}
\usepackage{graphicx}
\usepackage{enumitem}
\usepackage{textcomp}
\usepackage{hyperref}
\usepackage{subfig}
\usepackage[skip=13pt]{caption}
\usepackage{caption,setspace}
\def\BibTeX{{\rm B\kern-.05em{\sc i\kern-.025em b}\kern-.08em
		T\kern-.1667em\lower.7ex\hbox{E}\kern-.125emX}}
	
\usepackage{multirow}

\begin{document}
	\history{Date of publication xxxx 00, 0000, date of current version xxxx 00, 0000.}
	\doi{10.1109/ACCESS.2018.DOI}
	
	\title{Blockchain and AI-based Solutions to Combat Coronavirus (COVID-19)-like Epidemics: A Survey}
	\author{\uppercase{Dinh C. Nguyen}\authorrefmark{1}, (Graduate Student Member, IEEE), \uppercase{Ming Ding}\authorrefmark{2}, (Senior Member, IEEE), 
		\uppercase {Pubudu N. Pathirana\authorrefmark{1}}, (Senior Member, IEEE), 
		\uppercase{Aruna Seneviratne}\authorrefmark{3}, (Senior Member, IEEE)}

	\address[1]{School of Engineering, Deakin University, Waurn Ponds, VIC 3216, Australia}
	\address[2]{Data61, CSIRO, Australia}
	\address[3]{School of Electrical Engineering and Telecommunications, University of New South Wales (UNSW), NSW, Australia}
	
	\markboth
	{Author \headeretal: Preparation of Papers for IEEE TRANSACTIONS and JOURNALS}
	{Author \headeretal: Preparation of Papers for IEEE TRANSACTIONS and JOURNALS}
	
	\corresp{Corresponding author: Dinh C. Nguyen (cdnguyen@deakin.edu.au)}
	\begin{abstract}
	The beginning of 2020 has seen the emergence of coronavirus outbreak caused by a novel virus called SARS-CoV-2. The sudden explosion and uncontrolled worldwide spread of COVID-19 show the limitations of existing healthcare systems in timely handling public health emergencies. In such contexts, innovative technologies such as blockchain and Artificial Intelligence (AI) have emerged as promising solutions for fighting coronavirus epidemic. In particular, blockchain can combat pandemics by enabling early detection of outbreaks, ensuring the ordering of medical data, and ensuring reliable medical supply chain during the outbreak tracing. Moreover, AI provides intelligent solutions for identifying symptoms caused by coronavirus for treatments and supporting drug manufacturing. Therefore, we  present an extensive survey on the use of blockchain and AI for combating COVID-19 epidemics. \textcolor{black}{First, we introduce a new conceptual architecture which integrates blockchain and AI for fighting COVID-19. Then, we survey the latest research efforts on the use of blockchain and AI for fighting COVID-19  in various applications. The newly emerging projects and use cases enabled by these technologies to deal with coronavirus pandemic are also presented. A case study is also provided using federated AI for COVID-19 detection.}  Finally, we point out challenges and future directions that motivate more research efforts to deal with future coronavirus-like epidemics.  
\end{abstract}

\begin{IEEEkeywords}
	Blockchain, Artificial Intelligence (AI), security, privacy, machine learning, deep learning, coronavirus (COVID-19), SARS-CoV-2, epidemic. 
\end{IEEEkeywords}

\titlepgskip=-15pt

\maketitle

\section{Introduction}

\begin{figure}
	\centering
	\includegraphics [width=0.98\linewidth]{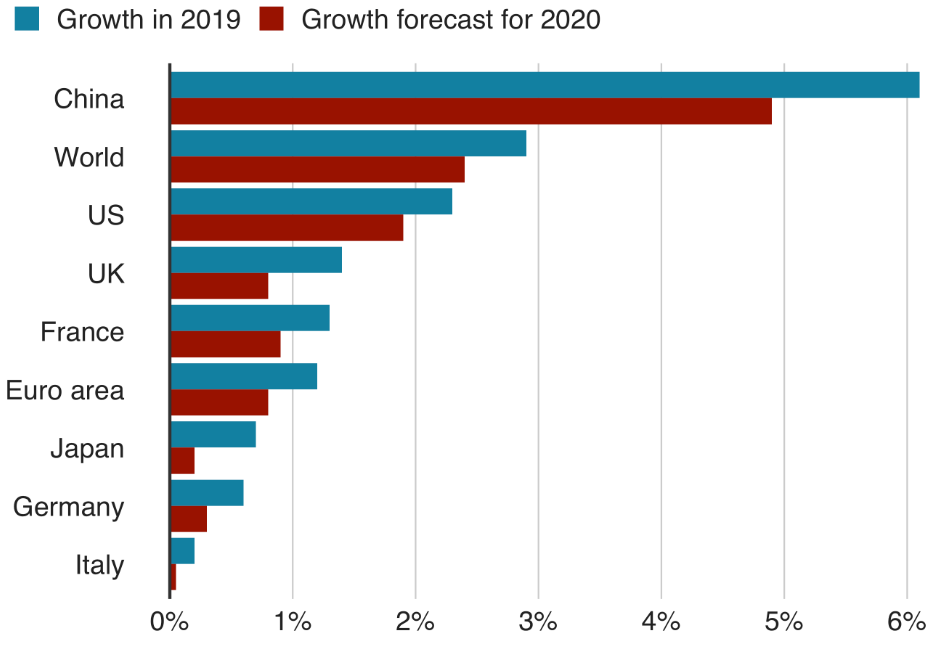}
	\caption{The impact of COVID-19 on global economic growth (Source: OECD). }
	\label{Fig:Chart}
\end{figure}

The coronavirus (COVID-19) outbreak in late 2019 comprises a serious threat around the world \cite{1}. The severity of the epidemic was so huge that the World Health Organization (WHO) was compelled to declare it as a pandemic within a month of its wide-scale expansion. The virus spread causes the global economic shock with the massive interruptions of many sectors such as supply chain, industry, insurance, agriculture, transport, and tourism, forcing governments and owners to shut stop operations on a worldwide scale \cite{2}.  According to the Organisation for Economic Cooperation and Development (OECD), the global economy could grow at its slowest rate since 2009 in this year \cite{3} due to the coronavirus outbreak as the forecasts in Fig.~\ref{Fig:Chart}. As the number of infections rises, many governments around the world have instituted drastic lock-downs and curfews and called for social distancing and work from home for virus spead fighting\cite{viet}.

With the deadly coronavirus spreading globally, every attempt is being made to ensure help for victims as well as to stop the spread. As governments scramble to address these problems, technology-empowered solutions can help deal with the worldwide health crisis. Applications of innovative technologies such as Blockchain and Artificial Intelligence (AI) could have the answers in response to coronavirus crisis \cite{4}, \cite{5}, \cite{6}. \textcolor{black}{Blockchain can support to combat the COVID-19 pandemic by enabling early detection of outbreaks, fast-tracking drug delivery, and providing a consensus on the ordering of the COVID-19 data records.} Furthermore, \textcolor{black}{ AI with supervised and unsupervised machine learning provides intelligent solutions for monitoring real-time epidemic outbreaks, pandemic trend forecasting, identifying symptoms caused by coronavirus for treatments, and supporting drug manufacturing \cite{411}.} It is also argued that the present coronavirus crisis should be seen as a "\textit{call to arms for the technology industry}" where blockchain and AI may be the key enablers for radically changing the landscape of crisis response and the management of the coronavirus outbreak \cite{7}, \cite{8}.  

\subsection{Motivations of Using Blockchain and AI for Coronavirus Fighting}
\subsubsection{	Limitations of Current Healthcare Systems}
The COVID-19 outbreak could push the existing healthcare systems to their limits. At present, there is a lack of a reliable data surveillance system that would instantly give relevant healthcare organizations the information they need about potential outbreaks. In fact, most of the current coronavirus information comes from separate sources such as the public, hospitals, clinical labs with a large amount of inaccurate data without being monitored thoroughly. The use of unreliable information makes it challenging for potential outbreak identification and quarantine. \textcolor{black}{Another limitation is the current time-consuming and in-accuracy coronavirus detection procedure that often takes several hours to complete the virus tests.} Thus, how to accelerate coronavirus detection speed with high accuracy is an urgent need. Furthermore, it is very challenging to process coronavirus data with complex patterns and large volumes by using human-depending medicine tools. 

\subsubsection{{Why we need blockchain and AI for fighting COVID-19 }}
{In this subsection, we will explain why there is a requirement of using blockchain and AI technologies for COVID-19 disease. More specifically, the reasons on why we need to use blockchain for such a disease are explained as follows. 
	\begin{itemize}
		\item \textcolor{black}{In such a pandemic scenario, blockchain provides promising security solutions to support  fighting COVID-19. Indeed, the blockchain constructs immutable ledgers of transactions for medical data sharing systems. Transactions (e.g., COVID-19 data) recorded in the blockchain cannot be modified or altered by any entities. Also, transactions are only written to blockchain while recovery actions or modifications are not allowed.  Moreover, all health data storage servers, e.g., cloud in the COVID-19 data collection, would be controlled autonomously by the blockchain and data usage over the network is always reflected on the blockchain for tracing \cite{dai2020blockchain}. More importantly, blockchain coupled with the smart contract technology eliminates the reliance on central servers to ensure fairness among transaction parties, e.g., patients, hospitals, governments and provide effective access control over data usage. Each entity in the blockchain-based COVID-19 data management system has an equal right to control all healthcare operations. Specially, blockchain with its distributed nature can work well when any party fails due to its decentralized nature \cite{garg2020anonymity}. Such traceability and decentralization are among the key unique features of blockchain that cannot be found in other traditional security techniques. }
		
		\item Moreover, blockchain can support well for reliable COVID-19 analytics. Collecting COVID-19 data is an important step of disease analytics. During the data collection, how to ensure the reliability of collected data is of practical significance to ensure the high quality of COVID-19 data analytics  [4]. The use of wrong data or unreliable database sources may result in biased analytical results, which can lead to fatal consequences like incorrect COVID-19 diagnosis. Moreover, in the emergency epidemic situation, many sources of COVID-19 data are collected from hospitals, the public, or media without being protected which can result in COVID-19 data modifications. These issues would definitely affect the correctness of the collected data which consequently reduces the reliability of the COVID-19 analysis process. In such contexts, blockchain is high of need to ensure the reliability of the collected data thanks to its security nature. \textcolor{black}{Blockchain also ensures the correct ordering of the COVID-19 data records  from the data sources to the destinations (e.g., hospitals or clinical labs) due to consensus mechanisms, which also ensures the high quality of data collection.} These features of blockchain would ensure correct data collection and thus support reliable COVID-19 analysis. 
		\item \textcolor{black}{Also, the incentive mechanisms offered by blockchain can be very useful to provide solutions for fighting COVID-19, by combining with crowdsourcing/crowdsensing technologies \cite{kadadha2020sensechain, more2020blockchain, kadadha2020abcrowd, lu2018zebralancer}. Indeed, crowdsourcing enables to use the crowd to collect information that is needed for the detection of COVID-19 patients. Besides, these frameworks incentivize patients and individuals to share information through payments, such as blockchain coins. For example, an artificial potential field-based incentive allocation mechanism is proposed in \cite{ lv2020towards} to incentivize IoT witnesses to pursue the maximum monitoring coverage and crowdsensing deployment in tracking human activities in the COVID pandemic using a decentralized and permissionless blockchain protocol. 
		\item Admittedly, the truthfulness of each medical data transaction is hard to verify since the involved medical entities might lie about the uploaded data. This make one wonder about the value of blockchain in fighting COVID-19, i.e., what is the use of a blockchain if it is full of lies? \footnote{It should be noted such an issue of fake data has been mitigated in our financial system thanks to the digital signature mechanism, e.g. verification will be performed for all of the transactions associated with network banking, online purchasing using debit/credit cards, etc.} Here, we argue that the immutability feature of blockchain would make the data fabrication significantly more difficult since the reported data shall be under scrutiny in the future by statistical and longitudinal data analysis. Hence, the true value of blockchain for combating COVID-19 is to immensely raise the difficulty and cost for the entities who dare to cheat and make up data.}
		
		
	\end{itemize}
	In addition, there are two key reasons on why we need to use AI for such a disease.
	\begin{itemize}
		\item In the epidemic, monitoring the trajectory of virus infection and estimating the potential outbreak are of paramount importance in the objective of controlling the COVID-19 disease. AI is able to learn from huge datasets on relevant factors including the number of cases, deaths, demographics, and related environments conditions. The learning capability of AI would help forecast future coronavirus cases and the size of a potential outbreak. It also helps build prediction models that can be used to estimate the outbreak in the future, which is very useful for governments in preparing necessary solutions in response to COVID-19. 
		\item 	To prevent the COVID-19 disease, modelling the coronavirus structure is very important to shed lights on which drugs could work against COVID-19. Due to the massive database collected from ubiquitous sources, developing a data learning and estimation system to estimate the COVID-19 virus structure is an urgent need that humans are unable to do without machine support. AI algorithms based on computer computation can learn data and predict how the proteins that make up an organism curve or crinkle based on their genomes to identify the shape of the receptor. This would facilitate the process of finding suitable drugs against COVID-19. 
\end{itemize}}
\subsubsection{Benefits of using blockchain and AI for solving the coronavirus epidemic}
Blockchain and AI are able to provide viable solutions to cope with the coronavirus epidemic from various aspects. \textcolor{black}{In fact, blockchain has been applied to other infectious epidemics, such as Ebola \cite{400} where blockchain was employed to conduct real-time Ebola contact tracing, transmission pattern surveillance and vaccine delivery. This project also reveals that the cryptographic feature of blockchain can help prevent unsecure data sharing between patients and health entities such as healthcare providers.} In the coronavirus outbreak fighting, blockchain can actively simplify the process of fast-tracking drug trials, and recording and tracking all fundraising activities and donations, in an immutable fashion, which can support the management of outbreaks and treatment.

\textcolor{black}{Moreover, AI has been applied to COVID-19-like epidemics. For example, AI techniques such as neural network architectures based on Long Short Term Memory (LSTM) were used to predict the influenza-like illness dynamics for military populations using neural networks and social media \cite{401}. Moreover, Bayesian ML was employed in \cite{402} to build predictive data learning models for identifying compounds active against the Ebola virus. These real-world use cases have demonstrated the potential of AI in supporting the fighting of infectious epidemics by forecasting the spread of virus and helping the drug production for treatment of infectious diseases.} In the coronavirus epidemic, AI can also provide solutions in several ways. AI can be used to detect virus and predict how the virus is going to spread by analysing the combined information of environmental conditions, access to healthcare, and the way it is transmitted. Based on that, AI can identify coronavirus within localized outbreaks of the disease and help reveal the nature of the virus \cite{10}. The coronavirus can cause severe symptoms such as pneumonia, severe acute respiratory syndrome, and kidney failure. AI-based algorithms such as genome-based neural networks already built for personalized treatment can prove very useful in controlling these adverse events or symptoms caused by a coronavirus, especially when the impact of the virus depends on immunity and the genome structure of individuals and no standard treatment can treat all symptoms effectively at present. \textcolor{black}{For example, ML-based regression models have been used in \cite{403} to estimate adverse outcome in COVID-19 pneumonia, by using chest CT scans with reverse-transcription polymerase chain reaction for severe acute respiratory syndrome coronavirus. More specifically, logistic regression models were used to evaluate the relationship between clinical parameters and CT metrics versus patient outcome (intensive care unit admission or death versus no ICU admission or death). The area under the receiver operating characteristic curve (AUC) was then computed to determine the adverse effects which support the clinical decision making for appropriate patient care.} Further, the use of AI may be very helpful in identifying the relationship of novel coronavirus and related viruses such as SARS \cite{11} to accelerate the finding of a new vaccine. Finally,  AI approaches can automatically build a model or relationship between treatments documented in healthcare records and the eventual patient outcomes. These models can quickly identify diagnosis and treatment choices that assist to guide the process of developing clinical guidelines for future coronavirus-like epidemics. With these promising benefits, recently the White House urgently calls for using AI to aid the US Government's response to the coronavirus pandemic \cite{12}. 
\subsection{Comparisons and Our Contributions}
\textcolor{black}{Recently, there are several recent research efforts made to survey the use of blockchain and AI for COVID-19 pandemic. The paper in \cite{survey1} focuses on analysing the roles of blockchain to address the crisis of trust during the COVID-19 pandemic. The authors in \cite{survey2} explain briefly the opportunities and limitations of the applications of blockchains in healthcare management, including COVID-19-related domains. The benefits of using blockchain for fighting COVID-19  are also explored in \cite{survey3}, with several specific use cases such as contact tracing, disaster relief, patient information sharing, e-government, supply chain management, online education, and  immigration management. }

Moreover, the roles of AI in fighting COVID-19 pandemic are also investigated recently. More specifically, the authors in \cite{survey5} explore the use of technologies including AI to help mitigate the impact of COVID-19 outbreak. The work in \cite{ting2020digital} provides a very brief discussion of the roles of DL for COVID-19 data analytics, along with a short introduction to blockchain in the pandemic. Another study in \cite{survey6} provides an extensive survey on the contributions of AI in combating COVID-19 from the aspects of disease detection and diagnosis, virology and pathogenesis, drug and vaccine development, and epidemic and transmission prediction. Furthermore, the use of popular AI techniques such as neural systems, classical SVM, and edge significant learning battling COVID-19 is analyzed in \cite{survey6}. Meanwhile, a survey paper in \cite{survey7} provides a systematic study of deep learning, deep transfer learning and edge computing to mitigate COVID-19. Other works in \cite{survey8, survey10} also survey the applications of AI techniques in screening, tracking, and predicting the spread of COVID-19. 

\begin{table*}
	\centering
	\caption{\textcolor{black}{Comparison our survey paper with existing works.  }}
	\label{Table:Comparisons}
	{\color{black}
		\setlength{\tabcolsep}{5pt}
		\begin{tabular}{|p{1cm}|p{2cm}|p{7.8cm}|p{5.5cm}|}
			\hline
			\centering \textbf{Related works}& 
			\centering \textbf{Topic}&	
			\textbf{Key contributions}&	
			\textbf{Limitations}
			\\ \hline
			\cite{survey1} &	Blockchain for trust in COVID-19 pandemic &	A discussion of the roles of blockchain to address the crisis of trust during the COVID-19 pandemic. &	The potential of blockchain for fighting COVID-19 such as secure outbreak monitoring, reliable supply chain  has not been investigated. 
			\\ \hline
			\cite{survey2} &	Blockchain in healthcare management &	A brief discussion of the opportunities and limitations of the applications of blockchains in healthcare management, including COVID-19-related domains. &	The analysis of blockchain usage for COVID-19 is very limited. 
			\\ \hline
			\cite{survey3} &	Blockchain for COVID-19 &	A survey on the use of blockchain for COVID-19 pandemic via several specific use cases. &	The potential of AI and its connection with blockchain for blockchain has not been discovered. 
			\\ \hline
			\cite{survey5} &	AI for COVID-19 &	A overview of the use of technologies including AI to help mitigate the impact of COVID-19 outbreak. &	The extensive survey on the blockchain for COVID-19 has not been provided. 
			\\ \hline
			\cite{ting2020digital} &	DL for COVID-19 &	A very brief discussion of the roles of DL for COVID-19 data analytics, along with a limited introduction to blockchain in the pandemic. &	The scope of the review of the DL and blockchain in COVID-19 is limited. 
			\\ \hline
			\cite{survey6} &	AI for COVID-19 &	A survey on the contributions of AI in combating COVID-19, from disease detection and diagnosis to transmission prediction. &	The extensive survey on the blockchain for COVID-19 has not been provided.
			\\ \hline
			\cite{survey7} &	DL and edge computing for COVID-19 &	provides a systematic study of deep learning, deep transfer learning and edge computing to mitigate COVID-19. &	The paper mostly focuses on computing aspects in the network edge for fighting COVID-19. 
			\\ \hline
			\cite{survey8} &	AI for COVID-19 prediction &	A survey of the applications of AI techniques in screening, tracking, and predicting the spread of COVID-19. &	The real benefits of AI for COVID-19 have not been verified. 
			\\ \hline
			\cite{survey10} &	AI techniques for COVID-19 &	A discussion of AI algorithms for COVID-19 data analytics. &	The analysis of the roles of AI for COVID-19 outbreak monitoring and prediction is missing. 
			\\ \hline
		\textit{Our paper} &Blockchain and AI for COVID-19 & A comprehensive survey on the applications of blockchain and AI for COVID-19. &-
			\\
			\hline
	\end{tabular}}
\end{table*}
{\color{black}Different from these survey works, our paper provides a comprehensive review on the use of both blockchain and AI for fighting COVID-19 pandemic. The key purpose of this paper is to provide the readers with an overall picture and road map of how blockchain and AI can support to solve the coronavirus epidemic and related healthcare crisis. Therefore, in this paper we present an extensive survey on the applications and use cases of blockchain and AI technologies specific to coronavirus (COVID-19) pandemic by using the rapidly emerging literature and the latest research reports. A case study is then provided using federated AI for COVID-19 detection. The unique challenges and future direction are also highlighted. The comparison of the related works and our paper is summarized in Table~\ref{Table:Comparisons}. To this end, our survey provides the following contributions:
	\begin{enumerate}
		\item We introduce a conceptual systematic architecture that integrates blockchain and AI for fighting COVID-19, aiming to provide the key solutions in response to coronavirus epidemic. 
		\item 	We identify a number of specific applications using these technologies for solving coronavirus-related issues detection. Particularly, we highlight the potentials of blockchain and AI through a comprehensive analysis and discussion in different applied scenarios.
		\item	We explore the latest use cases and projects using blockchain and AI for coronavirus fighting. 
		\item A case study is then provided using federated AI for COVID-19 detection, aiming  to demonstrate the benefits of the discussed technique in the pandemic. 
		\item	Based on the extensive survey, we identify possible research challenges and future directions to encourage scientists and stakeholders to put more efforts in developing innovative solutions to combat the future coronavirus-like epidemics. 
\end{enumerate} }
\textcolor{black}{\subsection{Methods and Materials}
	A systematic mapping study \cite{yli2016current} was selected as the research method for this survey paper, aiming to provide an overview of the research related to the use of blockchain and AI for fighting the COVID-19 pandemic, including the following key steps. First, we recognize the limitations of current healthcare systems and highlight the motivations on why we should use blockchain and AI for fighting COVID-19. The second stage is to search for all the relevant scientific papers on the research topic. We map the papers related to technical aspects of blockchain and AI for in supporting activities for COVID-19 pandemic prevention. Here we use the terms such as blockchain, machine learning, deep learning, COVID-19 as the search strings so that we can filter and select the most relevant technical papers for our survey. Then we chose the scientific databases for the searches. We decided to focus on peer-reviewed, high-quality papers published in conferences, workshops, symposiums, books and journals related to the research topic. We used four scientific databases for paper retrieval, including (1) IEEE Xplore, (2) ACM Digital Library, (3) Springer Link, and (4) ScienceDirect. The third stage is to screen all related papers based on their titles. Meanwhile, we exclude papers without high content quality, papers without text availability and papers that were not written by English. The fourth stage is keywording. We read the abstract and identified keywords and concepts that reflected the contribution of the paper. Then, we used the keywords to cluster and form categories for the mapping of the studies. The final stage is data extraction that gathers all information needed to survey technical aspects and contributions of bloclchain and AI in the context of COVID-19. }

\subsection{	Organization}
\textcolor{black}{The structure of this survey is organized as follows. Section II presents the background of coronavirus epidemic, and two key technologies including blockchain and AI, along with a conceptual architecture that integrates blockchain and AI for coronavirus fighting. In Section III, we analyse the solutions using blockchain for coronavirus fighting. Next, in Section IV, we present a review on the use of AI for coronavirus epidemic fighting. Section V summarizes the most popular use cases on blockchain and AI adaption in response to coronavirus fighting. We present a case study toward using federated AI for COVID-19 to demonstrate the feasibility of our approach in Section VI. Then, we point out several important challenges and potential future directions in Section VII for applying these technologies in COVID-19 epidemic. Finally, Section VIII concludes the paper. A list of key acronyms and abbreviations used throughout the paper is given in Table~\ref{Table:Acronyms}.}

\begin{table}
	\caption{{\color{black}List of key acronyms.}}
	\label{Table:Acronyms}
	\scriptsize
	\centering
	\captionsetup{font=scriptsize}
	\setlength{\tabcolsep}{5pt}
	{\color{black}\begin{tabular}{p{1.5cm}|p{3.6cm}}
		\hline
		\textbf{Acronyms}& 
		\textbf{Definitions}
		\\
		\hline
		COVID-19 &	Coronavirus
		\\
		WHO &	World Health Organization 
		\\
		CDC &	Centres for Disease Control
		\\
		IoT &	Internet of Things 
		\\
		AI& Artificial Intelligence
		\\
		ML & Machine Learning
		\\
		DL & Deep Learning
		\\
		FL& Federated Learning
		\\
		DNN &Deep Neural Network
		\\
		CNN & Convolutional Neural Network
		\\
		GAN & Generative Adversarial Network 
		\\
		CT & Computed Tomography 
		\\
		\hline
	\end{tabular}}
	\label{tab1}
\end{table}

\section{	Background}
In this section, we provide briefly the coronavirus pandemic, and then summarize blockchain and AI technologies used for fighting coronavirus. Then we introduce a systematic architecture for coronavirus fighting. 
\subsection{Coronavirus (COVID-19) Pandemic}
The first outbreak of the COVID-19 virus epidemic took place in Wuhan, a city of 11 million, the capital of Hubei Province, China, starting in December 2019 \cite{14}, {\cite{add3}}. It is not yet clear which animal the new coronavirus is associated with, although several data suggests it is similar to those found in bats. Coronaviruses affect the breathing passages, or respiratory tract. The number of infected citizens grew exponentially until Chinese authorities declared a complete lockdown of the affected region, the 60-million-people Hubei province. Symptoms can be mild like the common cold, including a runny nose and cough or they can be more severe and can cause difficulty in breathing. Other common symptoms includes fever, cough, and feeling tired. A few people become seriously ill but this seems more likely in the elderly and those with health problems \cite{15}. \textcolor{black}{According to the statistics reported by WHO on 21 September 2020 \cite{404}, there have been 30,949,804 confirmed cases of COVID-19, including 959,116 deaths. The COVID-19 disease has been spreading rapidly around the world, especially in the US, Southeast Asia and Europe.} The COVID-19 epidemic results in great harm to people's daily life and country's economic development. Governments have locked down cities, restricted movements of millions and suspended business operations that will slow down the global economy in the coming years. Undoubtedly, the virus outbreak has become one of the biggest threats to the global economy and financial markets \cite{18}, \cite{19}.
\subsection{Blockchain}
\begin{figure}
	\centering
	\includegraphics[width=0.98\linewidth]{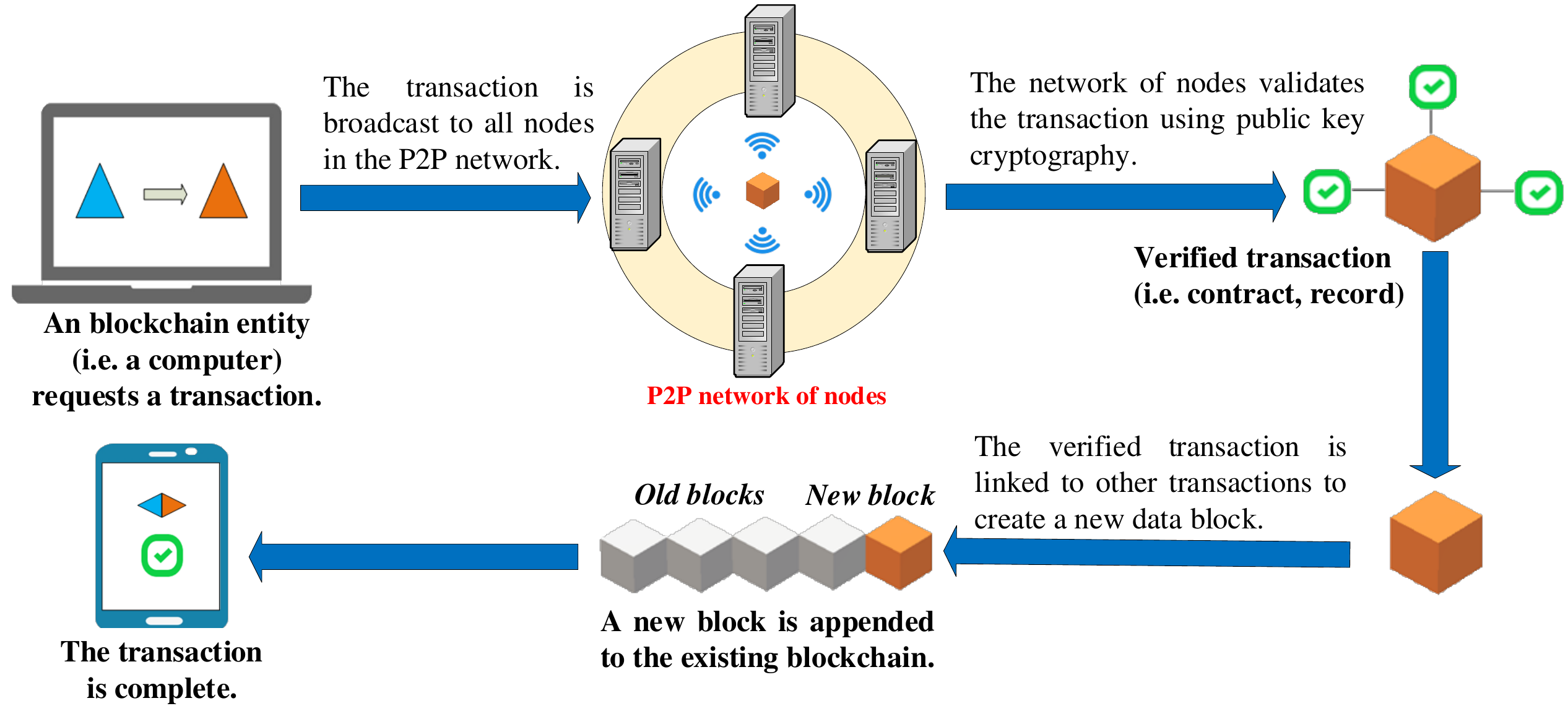}
	\caption{The concept of blockchain operation.}
	\label{Fig:Blockchain}
	\vspace{-0.1in}
\end{figure}

Blockchain is mostly known as the technological platform behind Bitcoin \cite{20}. The key concept of a blockchain is decentralization. In fact, its database does not place in a central location, but is distributed across a network of participants (e.g., computers). This decentralized concept provides high robustness and security for database stored on blockchain with no single-point failure. Importantly, blockchain is visible to each member in the network. This is enabled by a mechanism called consensus which is a set of rules to ensure the agreement among all participants on the status of the blockchain ledger. A concept of how blockchain works is shown in Fig.~\ref{Fig:Blockchain}. In general, blockchains can be classified as either a public (permission-less) or a private (permissioned) blockchain \cite{21}. A public blockchain is accessible for everyone and anyone can join and make transactions as well as participate in the consensus process. The best-known public blockchain applications include Bitcoin and Ethereum \cite{22}. Private blockchains on the other hand are an invitation-only network managed by a central entity. A participant has to be permissioned using a validation mechanism. Each blockchain consists of three main components, including data block, distributed ledger (database), and consensus algorithms. 

\textcolor{black}{Each data record (e.g., medical data) in the blockchain can be stored in the form of a transaction. Multiple transactions form a block, and multiple blocks are linked together to form a blockchain \cite{23}. The header field of each block contains the hash of the previous block, thus creating an ordered chain. The unique advantage of blockchain is its ability to ensure the chronological order of data records such as COVID-19 data. These data records are permanently stored on the chain in a chronological order. The consensus mechanism such as Proof of Work (PoW) in the blockchain ensures that the state of the entire blockchain is consistent and secure without the need for any third parties.} In the distributed ledger, each record contains a unique cryptographic signature decoupled with a timestamp which makes the ledger resistant to be modified. In terms of consensus algorithms, the process of transacting the block over the chain should not be controlled by any entity so that each block is managed by all participants with equal rights to avoid security issues, e.g., double-spending attacks. This can be done by a mechanism called consensus. From the view of blockchain, the consensus process mainly provides a guarantee for agreement on each data blockchain among entities. For instance, Bitcoin uses PoW algorithm \cite{23} as a main consensus scheme for its transaction management. The nodes with high computational capability can join the mining process and compete with each other to be a first one to verify the block. In return, the winner can receive a certain amount of coin as the reward for its mining effort. Along with the advance of blockchain, there are several new consensus algorithms such as Proof-of-stake (PoS), Byzantine Faulty Tolerant (BFT) \cite{24}. 

{In the blockchain ecosystem, a smart contract is an innovative technology for many applications, including healthcare \cite{sc2}. A smart contract is a programmable application that runs on a blockchain network. The term "smart contract" was defined by Nick Szabo as \textit{"a computerized transaction protocol that executes the terms of a contract. The general objectives of smart contract design are to satisfy common contractual conditions (such as payment terms, liens, confidentiality, and even enforcement), minimize exceptions both malicious and accidental, and minimize the need for trusted intermediaries"}. These condition scripts are placed on blockchain with a unique address. One can trigger a smart contract by sending a transaction to it. Due to the elimination of the involvement of external authorities, smart contracts potentially provide strong transparency for the involved blockchain network. Blockchain nodes within the network can execute the smart contract and derive the same results from the execution, which are reflected and recorded on the blockchain.} 

{Blockchain can connect with the Internet of Things (IoT) \cite{IoT1} in the context of healthcare for biomedical operations. In fact, IoT devices like sensors, gateways can collect real-time sensory data from patients and submit this data to a blockchain for sharing and storage in a secure manner. Specially, blockchain enables decentralized communication between IoT devices and IoT users (e.g., doctors, patients in healthcare) without the need of any central authority. Whenever an IoT activity (e.g., sensor data collection) is engaged, the blockchain nodes such as users in the network can monitor and agree on the transaction conditions and control the network operations. This process can be updated on the public ledger which is accessible to all network members while the role of central authority is eliminated \cite{IoT2}.} Up to date, blockchain has proved its success in healthcare and biomedical applications with promising performances in terms of healthcare data security \cite{25}, secure data management \cite{26}, \cite{27}, and transparent medical data storage \cite{28}, \cite{29}. Therefore, it is possible to apply in solving healthcare issues related to coronavirus epidemic. 
\subsection{Artificial Intelligence (AI)}
\begin{figure*}
	\centering
	\includegraphics[width=0.98\linewidth]{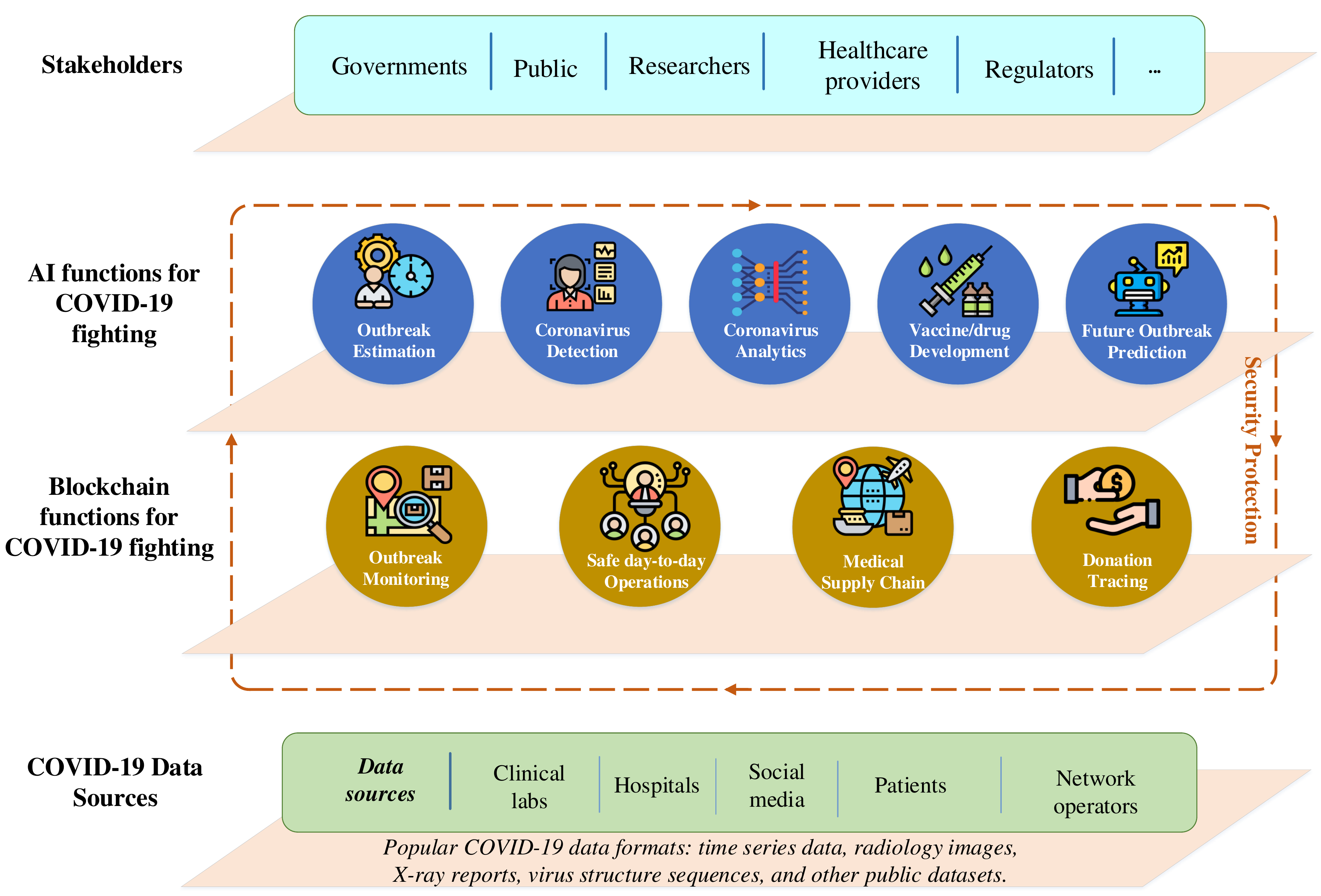}
	\caption{{Blockchain and AI for coronavirus fighting. } }
	\vspace{-0.1in}
	\label{Fig:Overview}
\end{figure*}
AI techniques have been used recently as a powerful tool for coronavirus data analytics,  prediction and drug/vaccine discovery  {\cite{add1}}, {\cite{add2}}. Recent studies show that AI has been mostly used for solving coronavirus-related issues via two key approaches: machine learning (ML) and deep learning (DL) \cite{qq}. ML is a subfield of AI. The objective of ML in general is to understand the structure of data and match that data into models that can be expressed and utilized by people \cite{31}. ML algorithms instead allow computers to perform training on data inputs and employ statistical analysis solutions to output values that fall within a specific range. Based on this, ML can provide building models from sample data to automate decision-making processes using data inputs. \textcolor{black}{For example, several preliminary studies in \cite{405} use temperature checking-based techniques such as facial and body temperature detection with smart helmet \cite{406} for detecting people with abnormal temperature, aiming to identify potentially infected persons.} For example, recently an AI company in the US specialising has utilized ML-powered interactive graphs are tracking the virus migration across China \cite{34}, for creating an alert system, whereby users will be able to receive information about whether an infected individual has travelled within their vicinity. This solution will help find infected individuals and provide medical resources to them. 

Meanwhile, AI methodologies such as DL have been also exploited to implement intelligent coronavirus fighting solutions. Conceptually, DL uses multiple neural network layers in a deep architecture \cite{35}. The neural network use multiple neurons connected via weighted connections among neural layers. For example, a basic deep neural network contains an input layer for receiving data samples, a single hidden layer for training and an output layer for generating training outcomes. Here, the number of hidden layers reflects the depth of the deep learning architecture. To generate the desired output, supervised or unsupervised learning techniques are used with the labeled or unlabeled data samples, associated with the adjustment of weight values among perceptrons \cite{36}. DL is known for AI applications such as speech recognition, computer vision, image processing, object detection for healthcare applications \cite{37}, \cite{38}, \cite{39} and thus coronavirus tackling can be a specific domain that DL can be applied for coronavirus fighting. Many AI companies have developed DL-enabled solutions to predict coronavirus infection and produce drugs and vaccines necessary for coronavirus disease \cite{40}. 

{\color{black}In the open literature, various AI-based approaches have been considered for COVID-19 detection, diagnosis and prediction. However, in the pandemic, collecting sufficient data to implement intelligent algorithms becomes more challenging and the user privacy concerns are growing due to the public data sharing with datacentres for COVID-19 data analytics.  In this context, federated learning (FL) \cite{FL0} has emerged as a promising AI solution for realizing cost-effective COVID-19 analytic-related applications with improved privacy protection. Conceptually, FL is a distributed AI approach which enables training of high-quality AI models by averaging local updates aggregated from multiple hospitals and medical centres without the need for direct access to the local data. This potentially prevents disclosing sensitive user information and user preference, and thus mitigates privacy leakage risks in this pandemic. Moreover, since FL attracts large computation and dataset resources from a number of COVID-19 data sources to train AI models, the COVID-19 data training quality, e.g., accuracy, would be significantly improved which might not be achieved by using centralized AI approaches with less data and limited computational capabilities.}
\subsection{Proposed Blockchain-AI Architecture for Coronavirus Fighting}
In this paper, we present an architecture as shown in Fig.~\ref{Fig:Overview} that integrates blockchain and AI for coronavirus fighting. The architecture is conceptually organized into four layers, including coronavirus data sources, blocchain functions, AI functions, and stakeholders. The work flow is explained as follows. 

\begin{figure}
	\centering
	\includegraphics[width=0.98\linewidth]{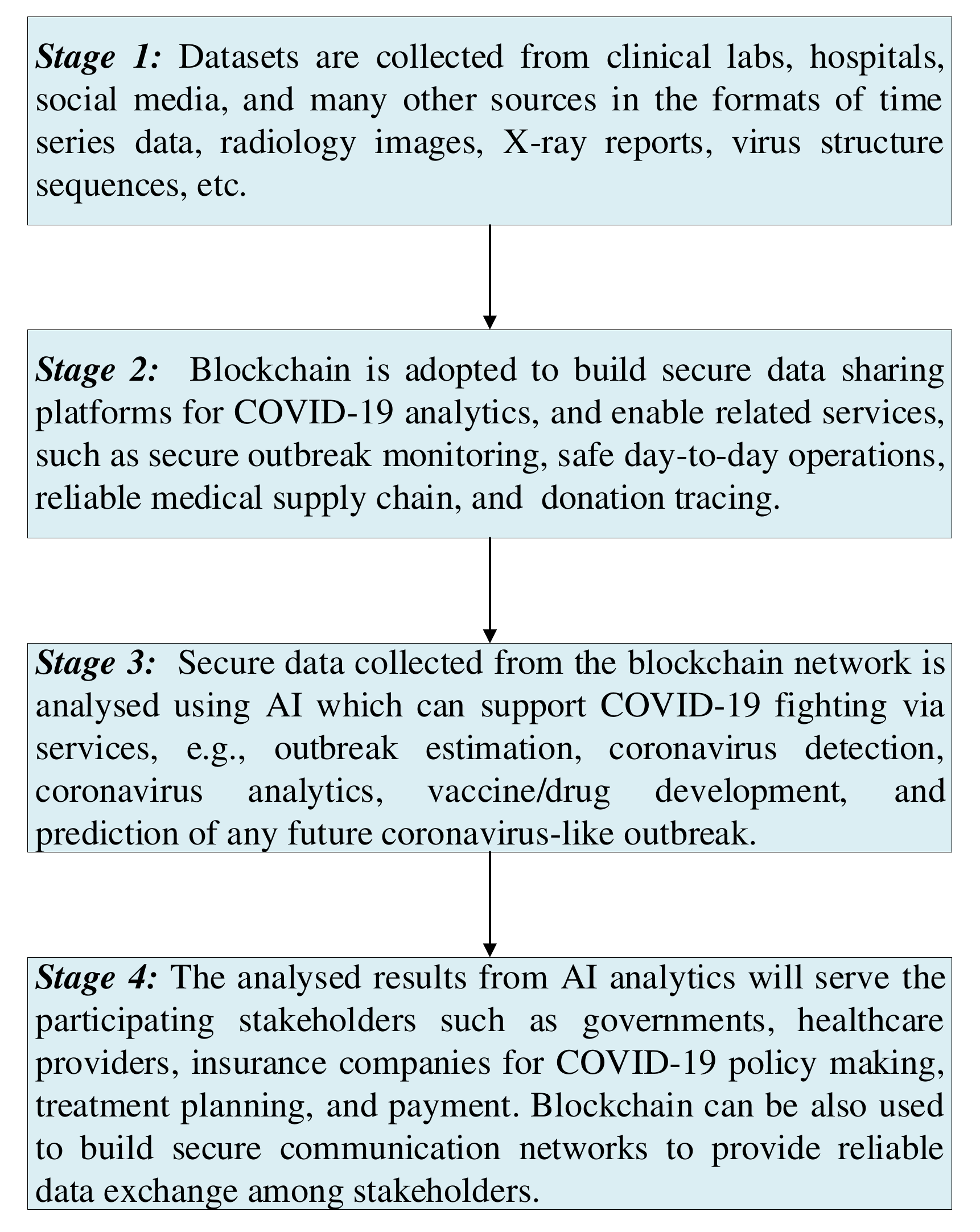}
	\caption{\textcolor{black}{The working flowchart of our proposed blockchain-AI architecture for fighting COVID-19. } }
	\vspace{-0.1in}
	\label{Fig:Flowchart}
\end{figure}

\begin{figure*}
	\centering
	\includegraphics[width=0.98\linewidth]{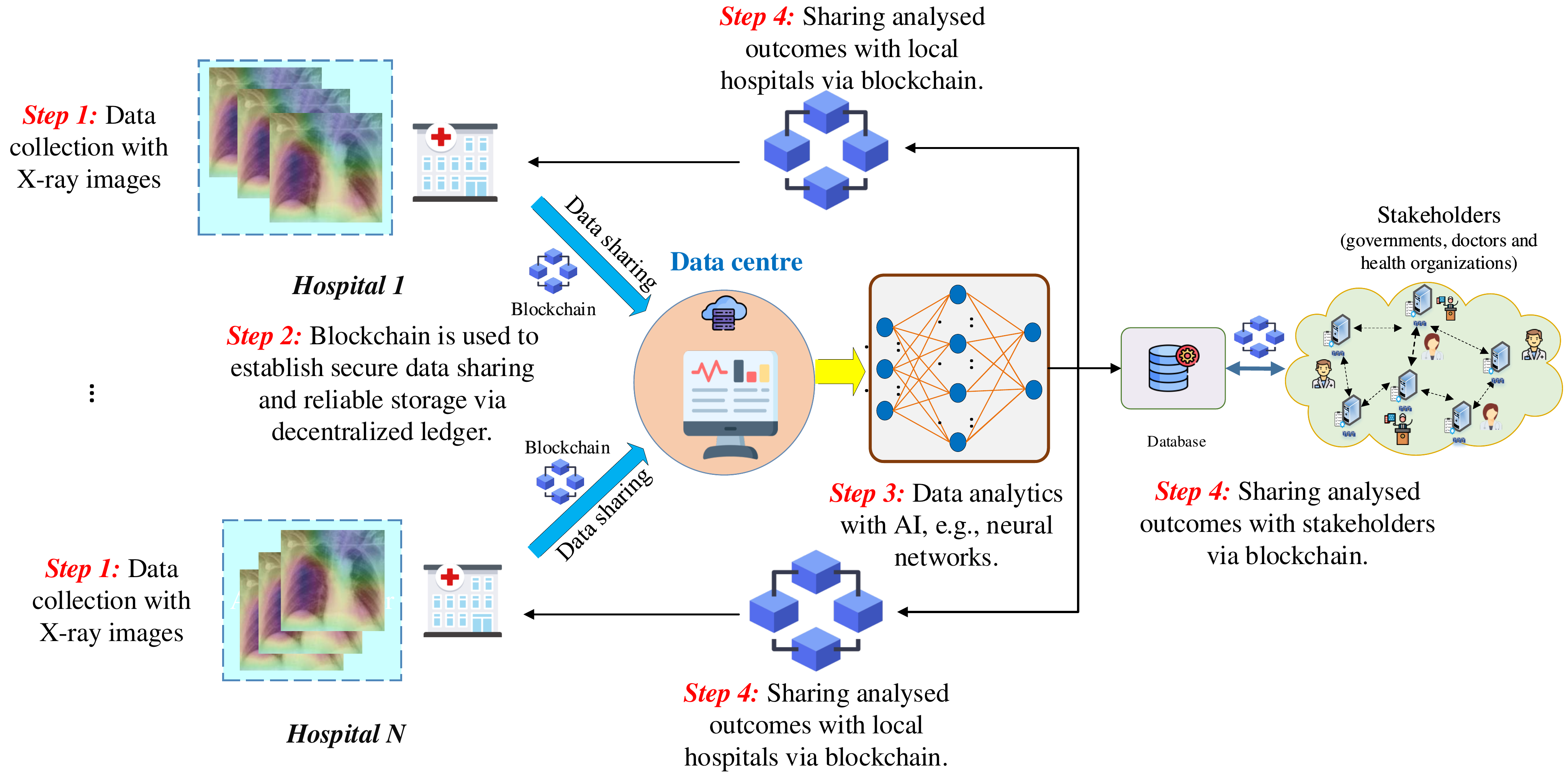}
	\caption{\textcolor{black}{A illustrative use case with blockchain and AI for COVID-19 data analytics. } }
	\vspace{-0.1in}
	\label{Fig:FLblockchain_Covid_casestudy}
\end{figure*}

{\color{black}Initially, all the data from clinical labs, hospitals, social media, and many other sources are consolidated and create raw data that subsequently develops in scale to big data. 
In the context of COVID-19, coronavirus data can include historic infectious cases, information of outbreak areas and coronavirus database in the formats of time series data, radiology images, X-ray reports and virus structure sequences that can be collected from various sources such as CDC, World Health Organization, clinical labs, the public and media. For example, recently China has built a large database to summarize the daily numbers of infected cases in time series updated from National Health commission Updates \cite{dinh1}.  The scientific community has also recently shared the database of chest X-ray or CT images on the open Github website for research usage \cite{dinh3}. These representative data sources would be valuable resources for both researchers and governments for the monitoring, analysis, estimation of COVID-19 epidemic. 

The data from these sources needs to be ensured for security during the coronavirus outbreak tracking and analytics, by using blockchain. Here, blockchain can offer a number of viable solutions for coronavirus-related services such as outbreak monitoring, safe day-to-day operations, medical supply chain, and donation tracing. Secure data collected from the blockchain network is analysed using intelligent AI-based solutions. By using reliable prediction and accurate analysis ability on big data collected from coronavirus sources, AI can provide support for coronavirus fighting via five main applications, namely outbreak estimation, coronavirus detection, coronavirus analytics, vaccine/drug development, and prediction of any future coronavirus-like outbreak. Finally, at the top of the hierarchy comes the stakeholder layer which includes parties such as governments, healthcare providers who are benefiting from blockchain-AI solutions. Note that blockchain can build secure communication networks and protocols to establish a fast and reliable data exchange with the stakeholders thanks to its decentralized nature \cite{41}. With this architecture, several blockchain-AI systems can be developed. For example, blockchain can establish the communication with hospitals and COVID-19 research centers and transfer securely COVID-19 data to the cloud computing where AI functions are located for data processing and computation.  Then, analyzed data can be returned via blockchain so that doctors physicians can do further steps for COVID-19 analytics \cite{mashamba2020blockchain}. The working flowchart of our proposed blockchain-AI architecture for fighting COVID-19  is summarized in Fig.~\ref{Fig:Flowchart}.  An illustrative use case from the proposed architecture is shown in Fig.~\ref{Fig:FLblockchain_Covid_casestudy}. To be clear, data collected from hospitals can be shared with the data centre via blockchain. In the data centre, AI tools can be employed, such as a neural network, for COVID-19 data analytics, e.g., classification, regression, etc.  Blockchain is also used to establish secure exchange of analyzed outcomes between the data centre, hospitals and the stakeholders via the decentralized ledger. }

\textcolor{black}{Note that the blockchain networks in COVID-19-related applications can include private and public ones. In the case of private blockchains, several solutions can be adopted to enhance security. For example, by introducing atomic meta-transactions proposed in \cite{ dorri2019spb}, blockchain entities can commit to their obligations during the data exchange (e.g., a private blockchain network between doctors and patients in a hospital). Particularly, smart contracts can be used to provide reliable authentication over the blockchain network where transactions and data exchange events can be verified in a transparent and reliable manner. In the case of public blockchains, secure data usage can be also accomplished according to the predefined access permissions of patients through the smart contracts of blockchain \cite{liu2018bpds}. Another solution is also proposed in \cite{ hasan2019combating} using a public Ethereum blockchain to prevent fake information, by enabling to trace and track the provenance and history of digital content to its original source even if the digital content is copied multiple times. The smart contract utilizes the hashes of the interplanetary file system (IPFS) used to store digital content and its metadata so that the content can be credibly traced to a trusted or reputable source. These proposed solutions are promising to improve the security and performance of blockchain systems for fighting COVID-19. }

In the following sections, we present an extensive survey on the latest research efforts on using blockchain and AI for coronavirus epidemic. 
\section{	Blockchain-based Solutions for Coronavirus Fighting}
In this section, we analyse the role of blockchain for coronavirus fighting via four key solutions, including outbreak monitoring, safe day-to-day operations, medical supply chain, and donation tracing, as shown in Fig.~\ref{Fig:Overview}. 
\subsection{Outbreak  Monitoring}
\textcolor{black}{Blockchain can provide feasible solutions for monitoring the coronavirus outbreak. Indeed, the blockchain can realize provide data visualization tools for deadly coronavirus monitoring. Blockchain can be seen as distributed database ledgers that can receive multiple updates in near real-time and store them on blocks linked together in a immutable manner. In the coronavirus epidemic, blockchain is potential to support the thousands of coronavirus victims by recording immutably patient symptoms of infection \cite{42}. This is highly important to monitor the spread of coronavirus because several patients can deliberately declare wrong their symptoms for avoiding hospital visits, which can lead to patient quarantine failure. Governments and healthcare organization can use blockchain to provide real-time data about affected areas and safe zones for providing fighting efforts \cite{43}. Information of safe zones such as population, location, current coronavirus outbreak status is recorded via a chain of blocks, and each block can store an update of the outbreak in a certain time. Note that such data can be collected from surveillance providers who use a combination of technologies like AI and geographical information systems (GIS). Based on that, healthcare service professionals can identify the free-virus zones from the infected ones to implement quarantine requirements \cite{45}. Blockchain hence can provide practical approaches to protect the community against virus spread. }

Another major problem in the coronavirus pandemic at this time is the outbreak of fake news circulated through social media channels and websites. We are facing an unprecedented crisis of public understanding. \textcolor{black}{In February 2020, Turkey and North Korea have been under scrutiny for claiming they have no diagnosed cases, but in fact the actual statistic from the WHO is different \cite{59}, while the accuracy on the case number given by Iran has been called into question, with lower numbers of infected cases and deaths.} Nowadays, people rely on social media platforms such as Facebook, Twitter to update the coronavirus pandemic, but these platforms are facilitators and multipliers of COVID-19-related misinformation. The incorrect information transmission really increases public confusion about who and what information sources to be reliable. The outbreak of fake news also generates fear and panic due to unverified rumours and exaggerated claims, and promotes racist forms of digital vigilantism and scapegoating \cite{60}. Governments, digital corporations and public health authorities should find solutions to verify and authenticate any information and news related to the pandemic which plays an important role in controlling public panic and facilitating coronavirus management. \textcolor{black}{Importantly, blockchain is able to provide a consensus on the ordering of the COVID-19 data records in the blockchain, aiming to achieve an agreement on the data among distributed entities such as healthcare providers, hospitals, and governments for better data management.}
\subsection{	Safe Day-to-day Operations}
\textcolor{black}{{In the coronavirus crisis, blockchain has emerged as a promising platform that enables to conduct day-to-day activities in virtual environments to reduce the risk of virus contraction.} In light of the ongoing coronavirus pandemic, customer services via virtual environments such as digital blockchain networks would become feasible to mitigate the spread of virus. For example, the UAE's Ministry of Community Development has switched to using digital channels via blockchain for civil services \cite{70}. Instead of visiting government offices and service centres for paper works, the government services such as the digital authentication of official certificates and other documents are implemented by blockchain. The trial results indicate that blockchain can process 2,919 different types of documents, which is even faster than the traditional working approaches. More importantly, this blockchain-based solution really helps to reduce the risk of infection from face-to-face contact that accordingly mitigates the possibility of virus spread in the community. }

\textcolor{black}{{Furthermore, economic activities can be continued using virtual platforms based on digital blockchain}. In fact, electronic payment among customers and companies now can be performed via blockchain instead of cash which may be a source of coronavirus spread \cite{53}. Blockchain is well known to make digital money such as Bitcoin \cite{54} which proves efficiency in certain digital transactions such as digital auction. In the coronavirus crisis that people should not use banknotes as Merchants are encouraging, electronic payment via blockchain may be an ideal choice. The decentralized concept of blockchain ensures payments in a secure, transparent, and immutable interface, which is free from human interaction. Another example is Ant Financial which has been launching a blockchain-enabled online bid opening system to replace the cash-based approaches \cite{55}. It allows customers to perform in contactless bidding from remote locations when they take quarantine measures to combat the COVID-19 outbreak. This bidding system is operated on a consortium blockchain platform which ensures that materials and processes of bid openings are tamper-proof, and the contactless bidding is transparent and trustworthy.}

\subsection{	Medical Supply Chain}
Blockchain has proved extremely useful in supply chain applications such as goods supply chain, trading supply chain \cite{62}, \cite{63}, \cite{64}, \cite{65}. In this pandemic crisis, maintaining a continuous supply of medicines and food has become a challenge for the healthcare sector. The blockchain technology can help the supply chain companies in achieving a fast flow of supply by tracking the flow from the origins to the destinations in a reliable manner. 

Recently, Alipay, along with the Zhejiang Provincial Health Commission and the Economy and Information Technology Department, China has launched a blockchain-based platform that allows users to track the demand and the supply chains of medical supplies \cite{66}. This consists of the recording and tracking of coronavirus epidemic fighting materials, such as masks, gloves and other protective gear. The company claims that blockchain can ensure high tracebility for the medical supply chain by secure linking among blocks and transactions and fast data flow thanks to blockchain decentralization.  When an outbreak occurs, a quick reaction and fast supply chain is the most important weapons authorities have to tackle the problem. By using blockchain, the supply chain issues can be solved, helping to save thousands of lives and billions of dollars \cite{67}. In summary, blockchain can offer five key solutions to support the medical supply chain in the coronavirus crisis:
\begin{itemize}
	\item 	Product requirements: provide a solution to update real-time demands and medical factories for fast response (e.g., supply chain rate adjustment).
	\item	Supply creditability: provide a solution to control the quality of goods from the factory side, e.g., product specification, supply volumes.
	\item	Transportation tracking: Goods, medical supplies need to be traced to guarantee transparency in the medical supply chain, which can be done by a blockchain network with transaction recording and monitoring capabilities.
	\item	Financial payments: blockchain can be used as a payment platform between the suppliers and the customers (e.g., users in quarantine areas). All digital payments are recorded on the blockchain with timestamp, signature without being changed or modified.
	\item	Customs certifications: customer behaviours such as buying, payment actions can be reflected on the blockchain with digitally signed certificates that shows the negotiation between the suppliers and the users. 
\end{itemize}
\subsection{	Donation Tracing}
The potential of blockchain in donation tracing applications has been investigated in recent works \cite{68}, \cite{69}. In the coronavirus crisis, donation is one of the most important activities to support livings and healthcare services for infected victims.  A critical question raised is how to trace the donation activities to ensure that donated goods, money are transferred to the targeted victims. Blockchain can be a feasible solution where the donation process can be traced through blockchain which issues signature and certificate to mark each of the donation updates, e.g., when, where, donation volumes, list of targeted receivers, etc. As an example, blockchain has been recently applied to trace the donation of food and protective gear like N95 masks to support citizens in the infected area \cite{71}. Blockchain also allows all parties such as suppliers, healthcare professionals, charity organizations to monitor the progress of donations. Any updates will be informed to all related parties to ensure transparency over the donation network. The Italian Red Cross is now also seeking Bitcoin donations to purchase highly needed medical equipment to infected areas \cite{72}. Binance Charity, a blockchain-based donation platform \cite{73}, also launches a million-dollars-campaign to seek donations via blockchain coins that are then exchanged to purchase supplies for supporting affected countries and regions. 

\begin{figure*}
	\centering
	\includegraphics[height=13cm, width=16.5cm]{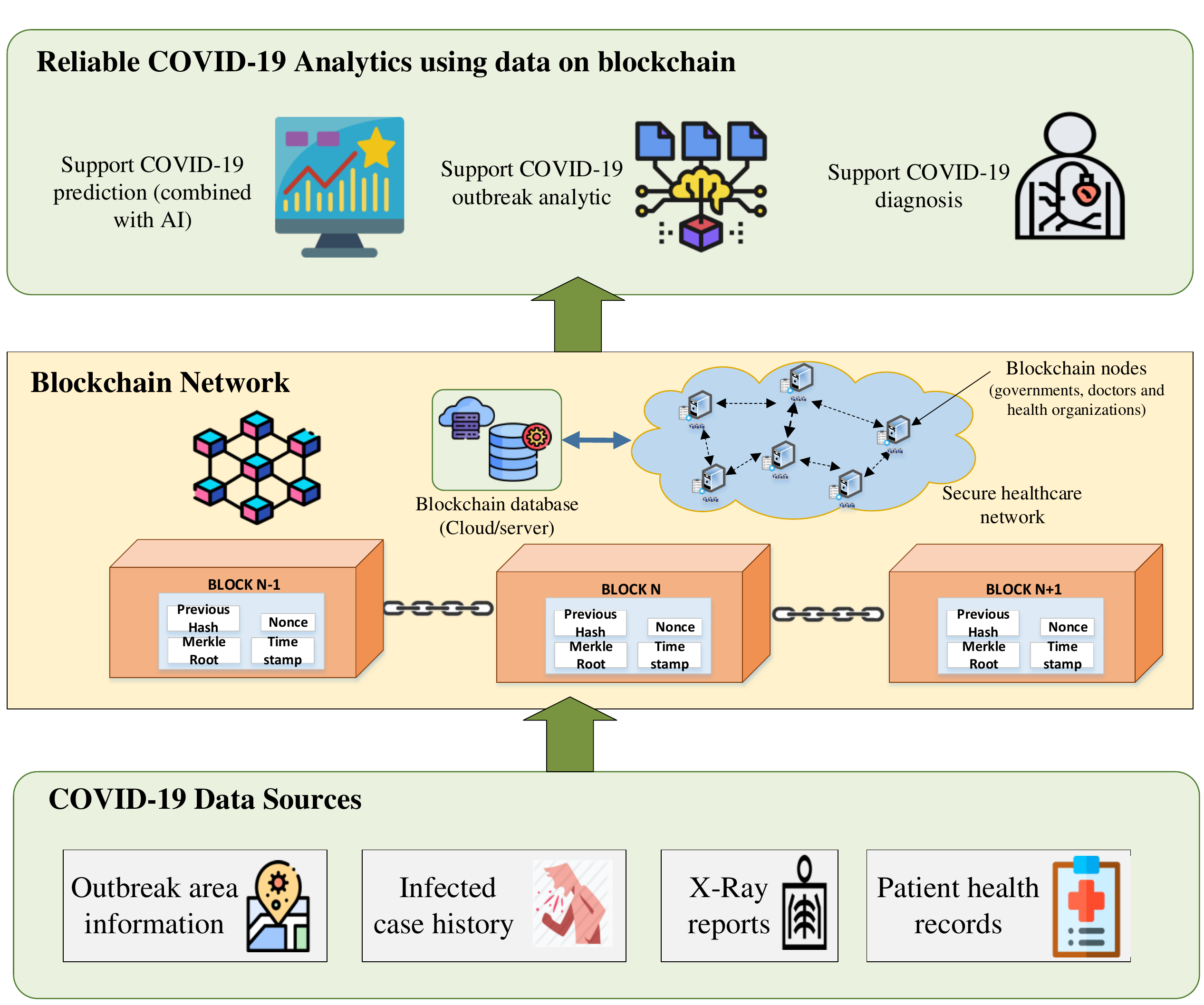}
	\caption{{The use case of blockchain for controlling COVID-19.}  }
	\label{Fig:BC_Casestudy}
\end{figure*}
{The Fig.~\ref{Fig:BC_Casestudy} presents a use case on how to integrate blockchain to support fighting COVID-19. COVID-19 data can be collected from different dataset sources such as outbreak area information, infected case history, X-ray reports and patient health records. Blockchain would build a database platform where COVID-19 data is attached with timestamps for data tracking in an immutable manner. In the COVID-19 context, many related parties are involved in a common blockchain ecosystem, such as governments, doctors, and healthcare organizations. Here, governments may need blockchain for epidemic tracking, and doctors want to collect reliably data for efficient COVID-19 prediction and diagnosis associated with the support of AI models (like ML, DL).  Moreover, healthcare organizations can also use blockchain to provide timely donations or medical supports to the patients. From our observation and discussions, the use of blockchain has the great potential to support strategies towards fighting the COVID-19 disease and control effectively the COVID-19 epidemic.} 
\section{	AI-based Solutions for Coronavirus Fighting}
In this section, we present a review on the use of AI for coronavirus epidemic fighting. AI can support by providing five main applications, namely outbreak estimation, coronavirus detection, coronavirus analytics, vaccine/drug development, and prediction of future coronavirus-like outbreak, as shown in Fig.~\ref{Fig:Overview}. 
\subsection{{Estimation of Coronavirus Outbreak Size}} 
{AI can help combat COVID-19 spread by its ability to estimate the coronavirus outbreak size through analysing people's phone usage patterns. In fact, when people are sick, dead due to coronavirus infection or taking care of family members, they tend to change their patterns of phone usage, which can be detected by AI algorithms run on the datasets provided by wireless operators. For example, if we can get access to the wireless operators' data in Wuhan from November 2018 to March 2020, then we may be able to find out the activity patterns of cell phones. 
	
	These patterns can include abnormal usage (e.g., frequent calls in early morning) from sick people, non-calling activities from people died from coronavirus infection, or new phone appearance in another city from people sneaked out of the lockdown city. Moreover, during home confinement in the lockdown period, sleep timing has markedly changed, with people going to bed and waking up later, and spending more time in bed \cite{407}. During outbreaks of infection, people are likely to experience fear of falling sick or dying themselves, feelings of helplessness, and stigma \cite{408}. \textcolor{black}{In addition to that, the upcoming economic downturn caused by COVID-19 crisis also leads to more consumption of unhealthy foods as they are cheaper. For example, according to a recent study \cite{food}, from December 2019 through February 2020 as the pandemic spread, the populations with a higher rate of poverty showed an increased consumption of unhealthy food. } In these situations, AI can analyse such different activities which can reveal the patterns of mobile users accordingly \cite{201}. For instance, ML is able to model and predict the personalized diverse activities of a user through learning from his phone usage records \cite{202}. Furthermore, DL would be a promising AI technique that relies on deep data learning and high-performance model prediction, aiming to estimate accurately the mobile application usage, e.g., abnormal calling behaviours, phone service inactivity \cite{203}. Therefore, such AI applications can be applied in the context of coronavirus epidemic to reveal the user patterns so that the agencies can estimate the size of coronavirus outbreak. }

{Moreover, a mobile operator can detect and track infected people's movement to predict the outbreak using AI analytics based on their cellular network location data. For example,  AI can analyse the user movement patterns to estimate the locations where crowds are still gathering and specify certain geographic quarantine areas. Meanwhile, stemming from the fact that people in several Chinese cities rely on their mobile phone for daily public transportation, Chinese railway authorities use AI to track the customers' movement and identify people who were near a passenger infected with COVID-19 \cite{200}. Human mobility patterns can be also estimated by using AI analytics based on geo-located Twitter social media data records \cite{205}. This solution can trace the geographical location of infected patients and keep track their movement to predict the coronavirus spread. }
\subsection{Coronavirus Detection}	
Several recently emerging solutions using AI for coronavirus detection have been proposed in the literature. A measure is to detect temperature on human face so that we can identify potential symptoms of COVID-19 disease. AI can come up with solutions thanks to its capability in face recognition applications. The work in \cite{74} presents an improvement in masked face detection using AI with real-world datasets  in response to the coronavirus epidemic. {Through deep learning, the dataset is trained to build an accurate masked face detection model, which then serves for the masked face recognition task, aiming to determine whether a person is wearing a mask.} Different from this study, the authors in \cite{75} rely on breathing characteristics to detect potential people infected with COVID-19. In the coronavirus detection, accurate identification of the unexpected abnormal respiratory pattern of people in a remote and unobtrusive manner is highly significant. Deep learning (DL) has been used for respiratory pattern classification, aiming to to classify six clinically significant respiratory patterns related to COVID-19 (Eupnea, Tachypnea, Bradypnea, Biots, Cheyne-Stokes and Central-Apnea). The work in \cite{78} uses image analysis-based techniques for coronavirus detection. An array of convolutional neural network based models (ResNet50, InceptionV3 and InceptionResNetV2) is leveraged to analyze the chest X-ray radiographs, aiming for detecting coronavirus pneumonia infected patients. The ResNet50 model can provide the highest classification performance with 98\% accuracy among the other two proposed models, as confirmed from simulations. 

To further improve the accuracy of COVID-19 detection, the authors in \cite{79} leverage a deep convolutional neural network that is able to automatically extract features starting from the genome sequence of the virus. Experiments using the 2019nCoVR dataset show a correct classification performance in detecting SARS-CoV-2, and distinguishing it from other coronavirus strains, such as MERS-CoV, SARS-CoV. Meanwhile, a research effort in \cite{80} is toward diagnostic uncertainty verification in the COVID-19 detection. A Bayesian Convolutional Neural Network (BCNN) is considered to estimate uncertainty in DL solutions. The objective of this study is to improve the diagnostic performance of the human-machine combination using a publicly available COVID-19 chest X-ray dataset. Another interesting work in \cite{81} presents a scheme for detecting coronavirus disease COVID-19 using on-board smartphone sensors. The proposed model enables us to read the signal measurements from smartphone sensors to identify viral pneumonia which can potentially identify coronavirus-related symptoms. It would be useful for radiologists with a smart phone in tracking the development of the disease. 
\subsection{Coronavirus Diagnosis and Treatment}
In the coronavirus fighting, developing innovative and efficient diagnostic and treatment methods has great significance, deciding the success of fighting COVID-19 disease. AI can come up with solutions to help understand and fight COVID19 via intelligent data analytics \cite{82}, \cite{83}. \textcolor{black}{A methodology is the thoracic CT image analysis to diagnose coronavirus positive patients \cite{76}. Non-contrast thoracic CT has been shown to be an effective tool in diagnosis, quantification and follow-up of corona-similar diseases. Given a potentially large number of thoracic CT exams, an analysis system is proposed that uses both robust 2D and 3D DL models coupled with existing AI models and combining them with clinical understanding. CT image analysis is also implemented in \cite{77} to early diagnose COVID-19 patients. The finding results indicate that the manifestations of computed tomography imaging of COVID-19 had their own characteristics, which are different from other types of viral pneumonia, such as Influenza-A viral pneumonia. Using these features, a 3D DL-based a location-attention classification model is developed using a pulmonary CT image set for symptom classification, showing a high accuracy with 86.7\% for three groups: COVID-19, Influenza-A viral pneumonia and healthy cases.} Moreover, the work in \cite{84} introduces the AI-based model, called COVIDX-Net that includes seven different architectures of deep convolutional neural network models, such as the modified Visual Geometry Group Network (VGG19) and the second version of Google MobileNet, for COVID-19 diagnosis. Each deep neural network model is able to analyze the normalized intensities of the X-ray image to recognize and classify the patient status either negative or positive COVID-19 case. Based on the current situation where there is a lack of public COVID-19 datasets, the learning-based scheme is validated on 50 Chest X-ray images with 25 confirmed positive COVID-19 cases, showing good performances in terms of high accuracy in COVID-19 classification. DL has been also used for the quantification of COVID-19 infection in CT images, as reported in \cite{85}. More specifically, a VB-Net neural network is employed to segment COVID-19 infection regions in CT scans, with a training dataset from 249 COVID- 19 patients, and then the model is validated using 300 new COVID-19 patients. The proposed scheme potentially assists radiologists to refine the automatic annotation of each infected case. 

\begin{figure*}
	\centering
	\includegraphics [height=6cm,width=13.8cm]{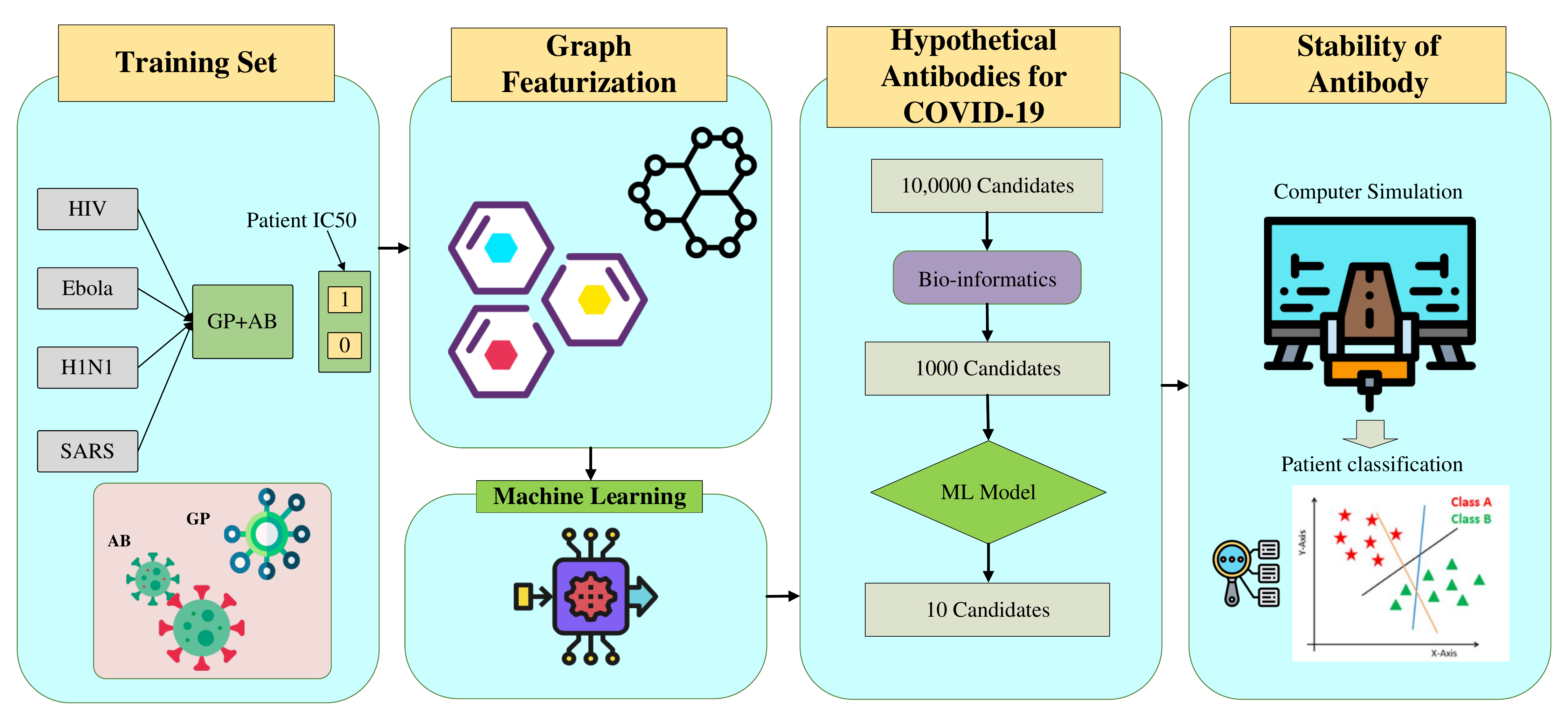}
	\caption{{An application of machine learning to design antibody against COVID-19 \cite{86}.} }
	\label{Fig:ML_Casestudy}
\end{figure*}

ML has been also utilized to facilitate coronavirus treatment. {The study in \cite{86} presents an example on using ML to design antibody for supporting COVID-19 treatment as shown in Fig.~\ref{Fig:ML_Casestudy}. In this work, the authors collects 1933 virus antibody sequences from clinical patient IC50 data. The graph featurization of antibody sequences has been integrated to provide molecular representation that is then trained by an ML model, such as a neural network, to generate thousands of potential candidates. Then, several ML classifiers such as random forest, support vector machine are used to classify these sequences to select the most optimal sequences and evaluate the stability of proposed antibodies. This workflow using ML would enable the rapid screening of potential antibodies that support the treatment of COVID-19 with high confidence. }

The work in \cite{87} also considers a based alignment-free approach for an ultra-fast, scalable, and accurate classification of whole COVID-19 genomes. Decision tree coupled with supervised learning has been also used for genome analysis from a large dataset of over 5000 unique viral genomic sequences. The authors claim that their scheme can achieve high levels of classification accuracy and discovers the most relevant relationships among over 5,000 viral genomes within a few minutes. Additionally, the research in \cite{88} suggests an ML-based miRNA prediction analysis for the SARS-CoV-2 genome to identify miRNA-like hairpin which can impact the COVID-19 virus. Learning can help to realize potential miRNA-based interactions between the viral miRNAs and human genes as well as human miRNAs and viral genes, which are important to identify the mechanisms behind the SARS-CoV-2 infections. \textcolor{black}{Furthermore, understanding personalized therapy is also an efficient approach for COVID-19 treatment \cite{409}. The remdesivir trials \cite{410} and preliminary data made available from recovery trial indicate that these therapies (antiviral and steroids, respectively) may have more benefit for certain patient populations and at certain timepoints within the illness. Understanding which patient and at what time would have maximum benefit from which cocktail of therapies would be clinically useful.}
\subsection{	Vaccine/drug Development }
In order to ultimately combat the emerging COVID-19 pandemic, it is desired to develop effective vaccines against deadly disease caused by COVID-19 virus. AI can come as an attractive tool to support vaccine production. A few rapidly emerging studies suggest to use AI for this task. For example, the work in \cite{89} employs Vaxign-ML tool to predict 24 COVID-19 vaccine candidates by using the S protein and five non-structural proteins that aims investigating the entire proteome of CoV-2. By applying reverse vaccinology and machine learning, the authors can predict potential vaccine targets for effective and safe COVID-19 vaccine development. Another vaccine trial is in \cite{90} which leverages HLA-binding prediction tools using an peptide stability assay. 777 peptides were assessed that were predicted to be good binders across 11 MHC allotypes with high prediction binding scores. Preliminary results can make important contributions to the design of an efficacious vaccine against COVID-19. 

Up to now, there is no effective treatment for this COVID-19 epidemic. However, the viral protease of a coronavirus can be extracted and replicated for drug discovery. The work in \cite{91} suggests that the potential anti-SARS-CoV chemotherapies can be potential 2019-nCoV drugs. A machine intelligence-based generative network complex model is derived to generate a family of potential 2019-nCoV drugs, inherited from HIV-1 protease \cite{92}. The results from this work can open up new opportunities to seek new drugs for fighting COVID-19 in this crisis. 

Unlike the previous work, the study in \cite{93} uses DL to develop a drug-target interaction model called Molecule Transformer-Drug Target Interaction (MT-DTI) to identify commercially available drugs that can act on viral proteins of 2019- nCoV. They suggest that atazanavir, an antiretroviral medication used for HIV fighting, can be useful to develop a chemical compound specific for COVID-19. The efficiency of the discovered drug is under the trial stage before possibly using on corona-infected patients. Moreover, the work in \cite{94} designs a data-driven drug repositioning framework, by integrating ML and statistical analysis approaches to systematically mine large-scale knowledge graph and transcriptome data. Results from trials show that ML can predict effective drug candidates against SARS-CoV-2. Meanwhile, drug repositioning is a possible solution for combating the global coronavirus crisis \cite{95}. A panel of 49 FDA-approved drugs that have been selected by an assay of SARS-CoV is monitored to recognize the potential drug candidates against COVID-19 infection. This trial drugs target to be tested on animals for tracking their antiviral activities before using on human. 
\subsection{	Prediction of Future COVID-19 Outbreak }
The recent emergence of coronavirus has highlighted the need for prediction of future outbreak like COVID-19. Epidemic models of COVID to predict and control the outbreak should be taken into consideration in the coronavirus fighting activities \cite{96}, \cite{97}, \cite{98}. Recently, AI has been applied to predict outbreak like coronavirus. For example, an prediction model using AI to estimate the size, lengths and ending time of Covid-19 across China \cite{99}. An auto-encoder is designed for modelling the transmission dynamics of the epidemic based on dataset collected from WHO sources. Clustering algorithms may be useful to classify different cities, regions for investigating the virus transmissions for outbreak prediction. Another model is developed for outbreak prediction in Mexico, as reported in \cite{100}. The biological features of the virus and the possible pathophysiological mechanisms of its disease and a stochastic model based on a Markov chain are used to express the probability distribution of cases in Mexico states. This also can estimate the number of cases in Mexico that facilitates the government in the estimation of virus spread making appropriate actions for future fightings. 

From the statistics collected from the COVID-19 outbreak, a research group in China built a prediction model for future infectious diseases using ML \cite{101}. The key idea is to extend causal inference theory and ML to identify and quantify the most important factors that cause zoonotic disease and COVID-19 outbreaks. Based on that, the model can generate visual tools to illustrate the complex causal relationships of animal infectious diseases and their correlation with COVID-19 diseases. Meanwhile, the authors in \cite{102} model the non-deterministic data distributions via DL network and fuzzy rule induction for obtaining better the stochastic insight about the epidemic development. Based on the model outputs, policy makers can have a better picture about the coronavirus spread and estimate the possibility of future similar outbreak. DL has been also tested in \cite{103} to build a virus host prediction model for identifying the way the virus transmits, with different sources such as bat coronaviruses, mink viruses and finding the relation with COVID-19. This research potentially discovers the origins of infection and predicts how coronavirus transmits for appropriate preventive actions. 

{\color{black}\subsection{Federated AI for Privacy-aware COVID-19 Data Analytics }
Many AI-based approaches have demonstrated significant promise in the intelligent COVID-19 data analytics for detection, diagnosis and prediction. However, in the pandemic, collecting sufficient data to implement intelligent algorithms becomes more challenging and the user privacy concerns are growing due to the public data sharing with datacentres for COVID-19 data analytics \cite{FL1}. FL can be an ideal candidate for assisting COVID-19 detection, by its federation and privacy protection features. In this way, each institution participates in training their DL model using their local COVID-19 images, e.g., X-ray and computed tomography (CT) images, and only model parameters such as gradients are exchanged while there is no need to share actual data and sensitive user information. For example, a number of hospitals can cooperatively communicate via the blockchain to run CNN updates locally for identifying CT scans of COVID- 19 patients \cite{FL2}. At each hospital, a deep capsule network is developed to enhance image classification performance, while FL provides guidance for transmissions of model updates to perform the model aggregation at a data centre. Simulations from 34,006 CT scan slices (images) of 89 subjects verify a high COVID-19 X-ray image classification and low data loss in the FL algorithm running. FL is also used in \cite{FL3}  to provide privacy-promoting AI solutions for COVID-19 chest X-ray image analytics. The preliminary experiments have been implemented, where multiple COVID- 19 X-ray image owners run a CNN-based model for image classification, and then share the computed parameters with a data centre for mobile averaging while the data ownership of each user is guaranteed. Four state-of-the-art CNN models, e.g., MobileNet, ResNet18, MoblieNet and COVID-Net, are used in the federated setting for evaluation, where ResNet18 is proven with the best COVID-19 detection performance (98.06\%) in federated X-ray image learning settings. 

Moreover, a dynamic fusion-based FL method is proposed in \cite{FL4} for CT scan image analysis to diagnose COVID-19 infections via two stages, including client participation and client selection. First, each client such as a medical institution make a decision to participate in the FL round based on the performance of the newly trained model. The central server also makes the decision to select which clients are permitted to update their local gradients by calculating the updating time. If a client cannot update its gradient within a predefined time interval, it is excluded from the FL aggregation. Another FL approach is designed in \cite{FL5}  for COVID-19 screening from Chest X-ray images by the collaboration of multiple medical institutions. The focus is on Chest X-ray images classification to identify COVID-19 from non-COVID-19 cases, where the feature extraction and the classification of X-ray images are performed based on a CNN to detect the COVID-19 disease. At each hospital, a CNN model is trained by allowing each X-ray image to be put into a convolutional layer and output the probability of COVID-19 infection, and then a central server is used to aggregate synchronously with the local institutions for building a strong classification model for COVID- 19 detection without compromising significantly user privacy which is valuable in the pandemic \cite{FL6}. Moreover, FL is also combined with DL to build a deep collaborative learning solution for detecting COVID-19 lung abnormalities in CT \cite{FL7}. The internal datasets are collected from a total of 75 patients confirmed COVID-19 infection at three local hospitals in Hong Kong for FL simulations, and then the generalizability is validated on external cohorts from Mainland China and Germany. }

\section{Blockchain and AI Use Cases for Coronavirus Fighting}
The use of blockchain and AI for solving coronavirus pandemic has been attracted much attention from industry and research community with projects and use case trials. In this section, we highlight the most popular use cases on blockchain and AI adaption in response to coronavirus fighting. 
\subsection{	Blockchain Use Cases}
\subsubsection{	Hashlog}
The first project is Hashlog that was created by a Georgia-based health tech startup Acoer \cite{104}. They has come up with a HashLog blockchain solution to combat and control the coronavirus from spreading. This can be enabled by the distributed blockchain ledger technology which ensures logging and data visualization of the coronavirus outbreak from the public data of US Centres for Disease Control (CDC) and WHO. Hashlog potentially provides real-time updates on the transmission of virus by tracking the movement of infected people which allows the healthcare authorities to make decisions for fighting further infections. 
\subsubsection{Hyperchain}	
The next blockchain project related to coronavirus is Hyperchain \cite{105}, which has built a donation tracking platform for supporting governments and healthcare organizations in the donation process to the infected victims in China. Hyperchain can ensure the transparency of the donation records from the origins to the destinations without being changed or modified. Moreover, the Hyperchain is able to connect up to millions of nodes so that more users can reach donated goods and necessary medical equipment from the factories, helping to solve the facility shortage issues during the epidemic. 
\subsubsection{	VeChain}
One interesting blockchain use case is VeChain which is a blockchain-based platform built for monitoring vaccine production in China \cite{106}. All activities related to vaccine manufacturing from materials, codes to package are recorded and stored on distributed ledgers. This project also provides a reliable method to reduce the risk of potential modifications on vaccine information. VeChain also ensures that vaccine records are immutable and permanent to achieve high vaccine quality which is highly important in the healthcare sector, like COVID-19 epidemic. 

\subsubsection{	PHBC}
Another promising blockchain platform is PHBC \cite{107} that is a monitoring blockchain for the continual and anonymous verification of communities and workplaces that are free from coronavirus COVID-19 as well as other high-risk viruses, bacteria, to aid them in staying free of deadly diseases. An interesting feature of this blockchain platform is the ability of monitoring the movement of uninfected persons, while restricting their return if they have gone to the areas known to be infected. PHBC can also automatically identify zones with and without validated incident reports by integrating the real-time information collected from virus surveillance providers with AI and geographical information systems. 
\subsection{AI Use Cases}
Various AI companies are also making a play towards efficient detection and diagnostic of diseases caused by COVID-19 virus. The important use cases and projects will be highlighted as follows. 
\subsubsection{Bluedot}
Bluedot is an AI project developed by a Toronto-based startup that aims to detect coronavirus in the epidemic \cite{108}. Bluedot use AI algorithms to build prediction models for virus ditection. It even identified the coronavirus outbreak in Wuhan hours before the local authorities had diagnosed the first cases of the virus. Bluedot collects information from social media, government documents, data of healthcare organizations to build intelligence enabled by natural language processing (NLP) and machine learning. It is able to track outbreaks of over 100 different diseases, every 15 minutes around the clock. 
\subsubsection{Infervision}
Another emerging AI use case in response to coronavirus pandemic is Infervision \cite{109} that aims to help clinicians detect and monitor the disease efficiently. Infervision has been used at the center of the epidemic outbreak at Tongji Hospital in Wuhan, China. Infervision AI has improved the CT diagnosis speed, which is highly important to accelerate pneumonia diagnosis for early detection of suspected patients.  This AI solution has proved very useful in supporting healthcare services in the emergency situations, like coronavirus outbreak. 
\subsubsection{AlphaFold system, Google DeepMind}
Very recently, Good DeepMind has introduced an AI tool called AlphaFold \cite{110} which enables the structure predictions of several under-studied proteins associated with SARS-CoV-2 virus that causes the coronavirus outbreak. This AI platform is empowered by a DL system and an ML-based free modelling technique to help it estimate protein structures when no similar structures of protein are available. Up to date, AlphaFold is under the trial stages and experiments have yet not been verified. 
\subsubsection{NVIDIA's AI}
Last month, a team of physicians at the Zhongnan Hospital, Wuhan China, used GPU-accelerated AI software called NVIDIA's AI, which was primarily used to detect cancer in lung CTs, to identify the visual signs of the COVID-19 virus \cite{111}. Data training was performed by using over 2,000 CT images from several of the first coronavirus patients in China, aiming to build AI models for detecting signs of pneumonia caused by coronavirus. This AI solution assist doctors in the virus recognition for fast diagnosis and treatment. 
{\color{black}\subsubsection{Federated AI}
Recently, a real-world federated AI project using FL for COVID-19 region segmentation in chest CT with the participation of medical institutions from China, Italy, and Japan is presented in \cite{FL8}. Specifically, a multi-national database consisting of 1704 scans from these three countries is collected for setting up the FL framework and compare with the traditional local training approach using 945 scans with the support of expert radiologists. Given the strict regulatory policy on data privacy, FL is proven as a promising solution for countries to collaborate in the COVID-19 segmentation and detection without worrying the user information leakage and lacks of dataset.}
\subsection{Combining Blockchain and AI for Combating COVID-19 }
{Recently, several efforts have been made to combine blockchain and AI technologies to build innovative solutions against the COVID-19 epidemic. For example, a research group in Spain leverages these two technologies to develop a blockchain and AI-based app that is able to predict the evolution of the COVID-19 pandemic \cite{combine1}. Blockchain has been used to ensure that people comply with the social distancing rules by creating digital identities and licenses that keep track of the daily activities of a person. Meanwhile, a deep learning architecture empowered by neural networks and rules-based symbolic systems is designed to build an estimation model by using the datasets of genetic profiles, medical records, and medical treatments. This AI-based model would help government officials to make appropriate decisions regarding social distancing and quarantine measures. }

{Moreover, the Immuno-oncology company that specialises in RNA therapeutics Mateon is recently cooperated with the startup Meridian IT to build a new drug manufacturing system for COVID-19 disease, by combining both blockchain and AI technologies \cite{combine2}. Due to regulatory concerns, much of current drug manufacturing faces many challenges with complex legacy processes and the manufacturing operations are largely human-involved.  These working procedures are less effective and potentially remains production errors due to human involvement. The new application is proposed by exploiting AI-based neural networks to automate the operating procedure for rapid drug production. Blockchain has been also used to ensure that the production line complies with the FDA regulations (the law set of Food, Drug, and Cosmetic Act policies). The combination of blockchain and AI thus would bring drug manufacturing to an advanced stage with a faster, more accurate, and more reliable working procedure that would enable timely drug supply for COVID-19 prevention.}

{\color{black}\section{A Case Study}
This section presents a case study towards the application of AI for COVID-19 detection, by  developing a federated AI approach using FL for COVID-19 classification. The details of our AI approach with FL will be provided as follows.


\begin{figure}
	\centering
	\includegraphics [width=0.99\linewidth]{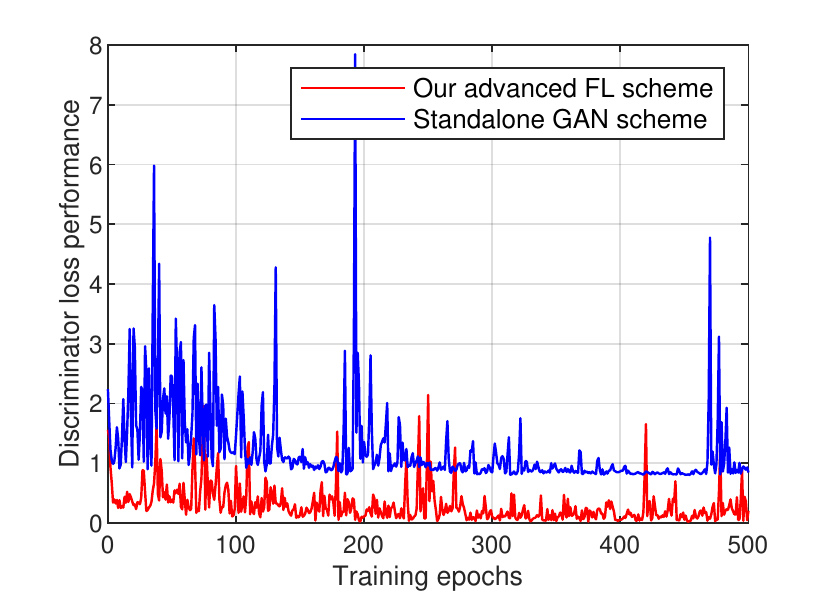}
	\caption{Performance of discriminator loss.}
	\label{Fig:convergence}
	\vspace{-0.1in}
\end{figure}

%

\subsection{System Model}
We consider a system model for the federated AI-based COVID-19 detection, including a set of medical institutions (e.g., hospitals) and a cloud server. Each institution participates in the FL process using its own COVID-19 X-ray image dataset, {e.g., X-ray images,} to build a global GAN with the cloud, aiming to generate high-quality synthetic COVID-19 X-ray images for improving the overall COVID-19 detection. Specifically, at each institution we design a GAN consisting of two components, namely a generator and a discriminator based on CNNs which alternatively train via a min-max game \cite{GAN1}. To implement the FL process, each institution participates in training its local GAN model by updating the parameters of the discriminator and the generator in each global round and exchanges them with the cloud server for parameter aggregation. For every global epoch, each institution collaboratively trains its discriminator and generator. First, the generator produces minibatches of fake samples from the noise probability distribution. Also, the discriminator samples minibatches of real data from the actual image distribution. Then, each institution simultaneously updates the discriminator and generator by ascending its stochastic gradients to update its own weights. After local training, all institutions transmit the learned updates to the cloud server for model averaging, while actual COVID-19 X-ray images are kept at local institutions which thus ensures data privacy. Then, the cloud server sends the new global update back to all participating institutions to prepare for the next round of FL training  \cite{FLdinh}. The learning process is iterated until the global loss function converges with a desired accuracy, where the synthetic COVID-19 X-ray images are generated with high-quality with respect to actual images. 
\subsection{Simulation Results}

\begin{table*}
	\scriptsize
	\centering
	{\color{black}
		\caption{Comparison of performance results for COVID-19 detection.}
		\begin{tabular}{|p{1.2cm}||ccc|ccc|ccc|ccc|}
			\hline
			\centering \multirow{2}{*}{\textbf{Classes}} &
			\multicolumn{3}{c|}{\textbf{Standalone scheme without GAN}} & \multicolumn{3}{c|}{\textbf{Standalone scheme with GAN}}  &
			\multicolumn{3}{c|}{\textbf{FL scheme without GAN}} & \multicolumn{3}{c|}{\textbf{Our advanced FL scheme}} \\ 		&Precision&	Sensitivity &	F1-score 
			&Precision&	Sensitivity &	F1-score
			&Precision&	Sensitivity &	F1-score
			&Precision&	Sensitivity &	F1-score
			\\ \hline
			COVID-19 &0.891 &0.865	&0.921  & 0.914&	0.872&	0.927&	0.950&	0.973&	0.986&	0.983&	0.993&	0.995
			\\  \hline
			Normal & 0.820&	0.996&	0.833&	0.928&	1&	0.980&	0.936&	0.981&	0.944&	0.946&	1&	0.962
			\\  \hline
			Pneumonia &	0.822&	0.448&	0.601&	0.912	&0.964&	0.963&	0.942&	0.796&	0.884&	0.964&	0.854&	0.918
			\\ \hline
		\end{tabular}
		\label{table:Performance_DarkCOVIDdataset}
		\vspace{-0.13in}}
\end{table*}
We report simulation results obtained when training a COVID-19 dataset \cite{11} of 620 X-ray images in three classes: COVID-19, normal, and pneumonia in an FL system with five institutions. By using the proposed FL model, we generate 1500 synthetic X-ray images which are then combined with an actual dataset for COVID-19 classification. We use a CNN-based classifier with three convolutional layers and Adam optimizer. We evaluate our approach and compare with the state-of-the-art schemes, including the standalone scheme (training dataset from an institution without federation), the standalone scheme with GAN \cite{18}, the FL scheme without GAN \cite{11}, and the centralized scheme (all datasets are transmitted to the cloud for classification). For reliable evaluation, the reported results are averaged from five runs of numerical simulations.

In Fig.~\ref{Fig:convergence}, {we compare the discriminator loss of our advanced FL scheme and the standalone scheme with GAN \cite{18}}, which only trains the GAN model using its dataset without the federation. It is observed that the standalone scheme is unable to achieve an optimal performance due to the lack of access to the full dataset which thus limits the quality of generated synthetic data. {Meanwhile, our advanced FL scheme} with GAN can learn over the entire data span from distributed institutions which achieves better minimum loss and helps extract better image features for efficient data augmentation.



We also evaluate the performance of COVID-19 detection via common quality metrics including precision, sensitivity, and F1-score. As illustrated in Table~\ref{table:Performance_DarkCOVIDdataset}, our advanced FL scheme outperforms all other state-of-the-art schemes in terms of these three metrics. For instance, for the COVID-19 class, the precision of our scheme is the best (0.983), while the standalone scheme without GAN, the GAN-based approach, and the FL scheme achieve lower performances, with 0.891, 0.914, and 0.950, respectively. Our scheme is also much better in sensitivity and F1-score performances. The intuition behind these observations is that our advanced FL scheme allows the cooperation of distributed institutions to enrich the global GAN model with a better COVID-19 X-ray image feature learning and a complete global COVID-19 X-ray image construction. As a result, our scheme can generate better realistic X-ray images that leads to higher detection rates. }

\section{	Challenges and Future Directions}
The application of blockchain and AI technologies is promising to help deal with the coronavirus epidemic. However, the provided throughout survey also reveals several critical challenges and open issues that should be considered carefully when applying these technologies in the context of healthcare sector for epidemic. We first highlight the research challenges, then we point out several potential future directions in this field. 
\subsection{	Challenges}
We here analyse the challenges from four main aspects: regulatory consideration, people's privacy preservation, security of blockchain and AI ecosystem, and the lack of unified databases. 
\subsubsection{Regulatory Consideration}
The use of blockchain and AI in the healthcare sector like coronavirus fighting should be considered carefully with regulatory laws. While the features of blockchain and AI can bring benefits, they also poses a legal and regulatory challenge if there is no party that is responsible and can be held accountable. For example, in the blockchain network, it will be important to consider what law might apply to transactions and what appropriate risk management should be put into place \cite{112}. Regarding AI, it might be easier to create several forms of legal schemes and internal governance models that will dictate the governing law for AI operations in healthcare. Specially, we also consider legal issues about content, personal information running on blockchain and AI platforms, such as problems with copyright infringement and defamation. 
\subsubsection{{People's Privacy Preservation}}
{In the coronavirus tracking applications, how to protect people's privacy is highly important. The governments can use mobile location data to help track the outbreak spread, but this solution must ensure the privacy of user data, especially sensitive information, such as home address, banking details, shopping records, etc. The governmental agencies may impose privacy laws on the user tracking mobile apps to ensure the safety and security to the public \cite{300}. Besides, nowadays many healthcare organizations and institutions are collecting data from their patients via electronic healthcare records that helps monitor the COVID-19 disease symptoms and serve treatment \cite{301}. In such healthcare activities, the conflict between data collection and user privacy is inevitable that needs to solved by laws and reinforcement from the related authorities. }
\subsubsection{	Security of Blockchain and AI Ecosystem}
Blockchain is widely regarded as secure monitoring platform to ensure safety and privacy healthcare applications, such as COVID-19 tracking. However, recent research reports have revealed inherent security weaknesses in blockchain which are related to medical and healthcare systems \cite{23}. Data threats or adversaries can enter the blockchain software to hold the control of the blockchain, which can lead to serious consequences, like modifications of medical transaction or patient data information, raising privacy concerns \cite{114}. Security is also a critical concern of AI systems for healthcare applications in the context of COVID-19 crisis. Data collected from patients, clinical labs and hospitals can be modified by data threats. More importantly, malicious attacks can inject false data or adversarial sample inputs which makes AI learning invalid, while the AI model can be tampered with \cite{115}. Therefore, security issues related to blockchain and AI deployment when solving the healthcare issues in the epidemic should be given high priority during the design and development stages. 
\subsubsection{Lack of Unified Databases}
A critical challenge in the coronavirus fighting is the lack of unified database related coronavirus epidemic such as infected cases, affected areas, and medical supply status. Most of the current coronavirus-related databases come from individual resources, e.g., social media \cite{116}, patient collection \cite{117}, \cite{118}, but they are not sufficient for large-scale AI operations that potentially creates greater impacts on fighting COVID-19, compared to the current results. Countries are now reluctant to share database, that make international healthcare organizations like WHO challenging to evaluate the epidemic for fighting effectively the virus spread. In fact, WHO cannot provide appropriate public health guidance without disaggregated data and detailed local epidemic information. 

\subsubsection{\textcolor{black}{Implementation Challenges}}
\textcolor{black}{Another challenge comes from the implementation of blockchain and AI platforms for COVID-19 prevention. In fact, the operation of blockchain requires powerful hardware and storage for running blockchains \cite{dev2014bitcoin}. Therefore, developing hardware specific to blockchain is vitally important to accelerate mining by a combined usage of computing elements within machines in blockchain networks, including private and public platforms such as Ethereum for COVID-19 tracing \cite{hasan2020blockchain}. Another critical issue is the extensive energy consumption and high network latency caused by running consensus processes such as PoW in the blockchain. This may hinder the applications of blockchain in distributed healthcare networks with resource-constrained IoT devices. Another problem is the limited throughput of blockchain systems. For example, Bitcoin can only process a maximum of four transactions/second, and the throughput of Ethereum achieved is about 20 transactions/second, while Visa can process up to 1667 transactions/second \cite{23}. Therefore, to make the applications of blockchain in COVID-19-related systems reliable, developing lightweight and throughput-efficient blockchain platforms are highly essential \cite{ sanka2018efficient}.}

\textcolor{black}{Moreover, the unprecedented increase of data volumes associated with advances of analytic techniques empowered from AI has promoted big data analysis, big data searching and big data mining activities for fighting COVID-19  \cite{costa2020meaningful}. The provision of big data platforms and their unified integration to the current COVID-19-related application of analytics, monitoring and research tools is thus highly needed.}

\subsection{	Future Directions}
We discuss several of the future research directions on blockchain and AI adoption for fighting COVID-19 pandemic.
{\color{black}\subsubsection{	Performance Improvement of Blockchain}
Blockchain platforms should be optimized to achieve better performances from different technical perspectives, such as scalability and network efficiency in terms of resource consumption, throughput, and network latency, aiming to make blockchain an ideal choice for emergency healthcare applications like the COVID-19 epidemic. For example, scalable and lightweight blockchain designs in healthcare are necessary to optimize data verification and transaction communication for ultralow-latency information broadcasting \cite{120}. In this regard, low-latency mining mechanisms with optimized block verification can be considered to mitigate delays in block processing for fast data sharing services that is necessary for efficient COVID-19 data analytics. Another feasible solution for blockchain performance enhancement is to minimize the size of blockchain by establishing local and private blockchain networks, one of these is responsible to monitor the outbreak in a certain area for fast response. To realize this, building customized ledgers that can placed on local servers in the outbreak area can help reduce latency in block broadcasting in the blockchain operations \cite{121}.}
{\color{black}\subsubsection{Security Issues of Blockchain}
Despite its great potential, blockchain still exists security issues, such as 51\% attack in the block mining and double spending attacks on data storage on blockchain. Therefore, innovative solutions should be considered for improving the security of blockchain. For example, a mining pool strategy is proposed in \cite{security1} to enhance the efficiency of the mining process, by solving security bottlenecks such as 51\% vulnerability with reduced block and transaction propagation delays. In this regard, defense mechanisms is integrated for fighting against data threats in the consensus process in the blockchain. Moreover, to mitigate double spending attacks in blockchain, a solution with recipient-oriented transaction processing is introduced in \cite{security2}, aiming to validate the transaction before appending to the block. This is enabled by using the concept of stealth address and master-node, where both transaction senders and recipients actively join the verification for each transaction in the propagation process. This allows for preventing double spending attacks using the verification time of the recipient and blocking time of the transaction.}
\subsubsection{Improved AI Algorithm for Better Analytic Accuracy}
The efficiency of intelligent healthcare data analytics, e.g., virus information analysis, mainly depends on the AI algorithms. Developing AI architectures specialized in medical applications might be the key for empowering future intelligent data analytics with the ability to handle multimedia healthcare data. Adaptive AI models should be developed in response to COVID-19-like emergency healthcare, such as AI in predictive modeling, AI in patient monitoring, and AI in emergency department operations \cite{122}. 

\subsubsection{Combination with Other Technologies} 
To achieve better efficiency in solving epidemic-related issues, blockchain and AI can be incorporated with other technologies to build a comprehensive healthcare system. For instance,  Alibaba has integrated AI with cloud computing for supporting coronavirus data analytics \cite{124}. The resourceful storage and high computation capability are the key features that cloud can provide to facilitate AI analytics. More interesting, China has recently exploited the mobility of drones \cite{125} to improve the provisions of medical supplies \cite{126}. In fact, using drones would be the fastest way to deliver essential facilities to the victims in the quarantine areas where all modes of transport are interrupted. Besides, drones also help to reinforce the contactless monitoring of the outbreak. In the near future, these promising technologies can be integrated to build highly advanced medical system for coping with the coronavirus-like epidemics. 
\section{Conclusions}
In this paper, we have presented an state-of-art survey on the utilization of blockchain and AI technologies to combat the coronavirus (COVID-19) epidemic. We have first introduced a conceptual architecture which integrates blockchain and AI towards the fighting of coronavirus crisis. Specially, we have extensively discussed the key roles of blockchain for solving the pandemic via five important solutions, including outbreak tracking, user privacy protection, safe day-to-day operations, medical supply chain, and donation tracking. Moreover, the potential of AI for coping with the COVID-19 crisis has been also analysed though five main application domains, namely  outbreak estimation, coronavirus detection, coronavirus analytics, vaccine/drug development, and prediction of future coronavirus-like outbreak. The important use cases and projects using blockchain and AI towards fighting COVID-19  have been also highlighted, followed by a case study. Finally, we have pointed out several potential challenges and future directions. We believe our timely survey will shed valuable light on the research of the blockchain and AI for fighting COVID-19  as well as motivate the interested researchers and stateholders to put more efforts into using these promising technologies to combat future coronavirus-like epidemics.


\bibliography{Ref1}
\bibliographystyle{IEEEtran}
\begin{IEEEbiography}[{\includegraphics[width=1in,height=1.25in,clip,keepaspectratio]{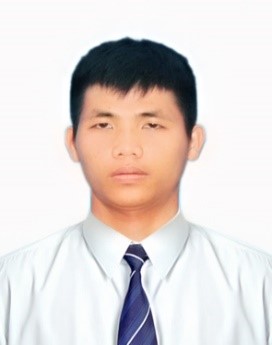}}]{Dinh C. Nguyen}
	(Graduate Student Member, IEEE) is currently pursuing the Ph.D. degree at the School of Engineering, Deakin University, Victoria, Australia. His research interests focus on blockchain, deep reinforcement learning, mobile edge/cloud computing, network security and privacy. He is currently working on blockchain and reinforcement learning for Internet of Things and 5G networks. He has been a recipient of the prestigious Data61 PhD scholarship, CSIRO, Australia.
\end{IEEEbiography}
\vskip -2\baselineskip 
\begin{IEEEbiography}[{\includegraphics[width=1in,height=1.25in,clip,keepaspectratio]{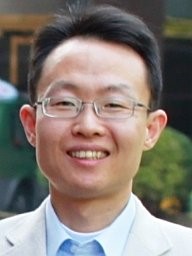}}]{Ming Ding}
	(M'12-SM'17) received the B.S. and M.S. degrees (with first-class Hons.) in electronics engineering from Shanghai Jiao Tong University (SJTU), Shanghai, China, and the Doctor of Philosophy (Ph.D.) degree in signal and information processing from SJTU, in 2004, 2007, and 2011, respectively. From April 2007 to September 2014, he worked at Sharp Laboratories of China in Shanghai, China as a Researcher/Senior Researcher/Principal Researcher. Currently, he is a senior research scientist at Data61, CSIRO, in Sydney, NSW, Australia. His research interests include information technology, data privacy and security, machine learning and AI, etc. He has authored over 140 papers in IEEE journals and conferences, all in recognized venues, and around 20 3GPP standardization contributions, as well as a Springer book "Multi-point Cooperative Communication Systems: Theory and Applications". Also, he holds 21 US patents and co-invented another 100+ patents on 4G/5G technologies in CN, JP, KR, EU, etc. Currently, he is an editor of IEEE Transactions on Wireless Communications and IEEE Wireless Communications Letters. Besides, he has served as Guest Editor/Co-Chair/Co-Tutor/TPC member for many IEEE top-tier journals/conferences and received several awards for his research work and professional services.
\end{IEEEbiography}
\vskip -2\baselineskip 
\begin{IEEEbiography}[{\includegraphics[width=1in,height=1.25in,clip,keepaspectratio]{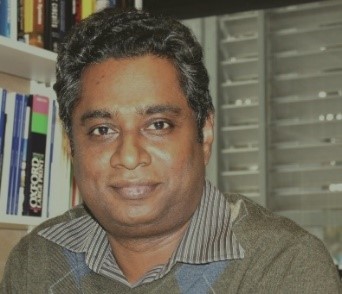}}]{Pubudu N. Pathirana}
	(Senior Member, IEEE) was born in 1970 in Matara, Sri Lanka, and was educated at Royal College Colombo. He received the B.E. degree (first class honors) in electrical engineering and the B.Sc. degree in mathematics in 1996, and the Ph.D. degree in electrical engineering in 2000 from the University of Western Australia, all sponsored by the government of Australia on EMSS and IPRS scholarships, respectively. He was a Postdoctoral Research Fellow at Oxford University, Oxford, a Research Fellow at the School of Electrical Engineering and Telecommunications, University of New South Wales, Sydney, Australia, and a Consultant to the Defence Science and Technology Organization (DSTO), Australia, in 2002. He was a visiting professor at Yale University in 2009. Currently, he is a full Professor and the Director of Networked Sensing and Control group at the School of Engineering, Deakin University, Geelong, Australia and his current research interests include bio-medical assistive device design, human motion capture, mobile/wireless networks, rehabilitation robotics and radar array signal processing.
\end{IEEEbiography}

\vskip -2\baselineskip 
\begin{IEEEbiography}[{\includegraphics[width=1in,height=1.25in,clip,keepaspectratio]{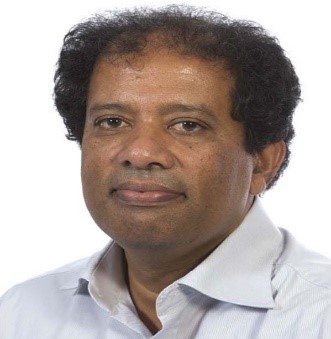}}]{Aruna Seneviratne}
	(Senior Member, IEEE) is currently a Foundation Professor of telecommunications with the University of New South Wales, Australia, where he holds the Mahanakorn Chair of telecommunications. He has also worked at a number of other Universities in Australia, U.K., and France, and industrial organizations, including Muirhead, Standard Telecommunication Labs, Avaya Labs, and Telecom Australia (Telstra). In addition, he has held visiting appointments at INRIA, France. His current research interests are in physical analytics: technologies that enable applications to interact intelligently and securely with their environment in real time. Most recently, his team has been working on using these technologies in behavioral biometrics, optimizing the performance of wearables, and the IoT system verification. He has been awarded a number of fellowships, including one at British Telecom and one at Telecom Australia Research Labs.
\end{IEEEbiography}

\EOD
\end{document}